\PassOptionsToPackage{pdfpagelabels=false}{hyperref} 
\documentclass{sigchi}

\toappear{\scriptsize Permission to make digital or hard copies of all or part of this work for personal or classroom use is granted without fee provided that copies are not made or distributed for profit or commercial advantage and that copies bear this notice and the full citation on the first page. Copyrights for components of this work owned by others than the author(s) must be honored. Abstracting with credit is permitted. To copy otherwise, or republish, to post on servers or to redistribute to lists, requires prior specific permission and/or a fee. Request permissions from permissions@acm.org. \\
{\emph{CHI '20, April 25--30, 2020, Honolulu, HI, USA.} } \\
\copyright~2020 Copyright is held by the owner/author(s). Publication rights licensed to ACM. \\
ACM ISBN 978-1-4503-6708-0/20/04\ ...\$15.00.\\
DOI: \url{https://dx.doi.org/10.1145/3313831.3376536} }

\usepackage{multirow}
\usepackage[caption=false]{subfig}

\usepackage{balance}       
\usepackage{graphics}      
\usepackage[T1]{fontenc}   
\usepackage{txfonts}
\usepackage{mathptmx}
\usepackage[pdflang={en-US},pdftex]{hyperref}
\usepackage{color}
\usepackage{tabularx}
\usepackage{booktabs}
\usepackage{textcomp}

\usepackage{microtype}        
\usepackage{ccicons}          

\usepackage{todonotes}

\def\plaintitle{Heartbeats in the Wild: A Field Study Exploring ECG Biometrics in Everyday Life}

\def\emptyauthor{}
\def\plainkeywords{Electrocardiogram; ECG; biometrics; field study}

\makeatletter
\def\url@leostyle{%
  \@ifundefined{selectfont}{
    \def\UrlFont{\sf}
  }{
    \def\UrlFont{\small\bf\ttfamily}
  }}
\makeatother
\def\pprw{8.5in}
\def\pprh{11in}

\definecolor{linkColor}{RGB}{6,125,233}
\hypersetup{%
  pdftitle={\plaintitle},
  pdfauthor={\emptyauthor},
  pdfkeywords={\plainkeywords},
  pdfdisplaydoctitle=true, 
  bookmarksnumbered,
  pdfstartview={FitH},
  colorlinks,
  citecolor=black,
  filecolor=black,
  linkcolor=black,
  urlcolor=linkColor,
  breaklinks=true,
  hypertexnames=false
}

\begin{document}

\title{\plaintitle}

\numberofauthors{2}
\author{%
  \alignauthor{Florian Lehmann\\
    \affaddr{LMU Munich}\\
    \affaddr{Munich, Germany}\\
    \email{research@florianlehmann.de}}\\
  \alignauthor{Daniel Buschek\\
    \affaddr{Research Group HCI + AI, Department of Computer Science, University of Bayreuth}\\
    \affaddr{Bayreuth, Germany}\\
    \email{daniel.buschek@uni-bayreuth.de}}
}

\maketitle

\begin{abstract}
This paper reports on an in-depth study of electrocardiogram (ECG) biometrics in everyday life. We collected ECG data from 20 people over a week, using a non-medical chest tracker. We evaluated user identification accuracy in several scenarios and observed equal error rates of 9.15\% to 21.91\%, heavily depending on 1) the number of days used for training, and 2) the number of heartbeats used per identification decision. We conclude that ECG biometrics can work in the wild but are less robust than expected based on the literature, highlighting that previous lab studies obtained highly optimistic results with regard to real life deployments. We explain this with noise due to changing body postures and states as well as interrupted measures. We conclude with implications for future research and the design of ECG biometrics systems for real world deployments, including critical reflections on privacy.
\end{abstract}




\section{Introduction} 

Specific human characteristics can be used to automatically identify a person and to subsequently grant access to a restricted area or to unlock devices. Several such biometric systems have already penetrated the market~\cite{Goode2014}. For example, fingerprint scanners or face recognition allow us to unlock our smartphones and notebooks. Moreover, biometrics can be used as an additional authentication factor to improve security~\cite{Khan2015}. 

Biometrics are also useful for access control in special working areas with a high demand for security, for example, laboratories, data centers, power plants, clean rooms, or hospitals. Here, biometric sensors are commonly applied to doors or panels near an entry point, for example, to recognise the user's fingerprint~\cite{Cole2006}, finger or palm veins~\cite{Zhou2011}, iris~\cite{Bowyer2008}, and face~\cite{Darwaish2014}. Behaviour~\cite{Pfeuffer2019} can also be used for identification, including voice~\cite{Campbell1997}, typing on a keyboard~\cite{Buschek2015chi}, or handwriting \cite{Tolosana2018}. 

\begin{figure}[t]
  \centering
  \includegraphics[width=1\columnwidth]{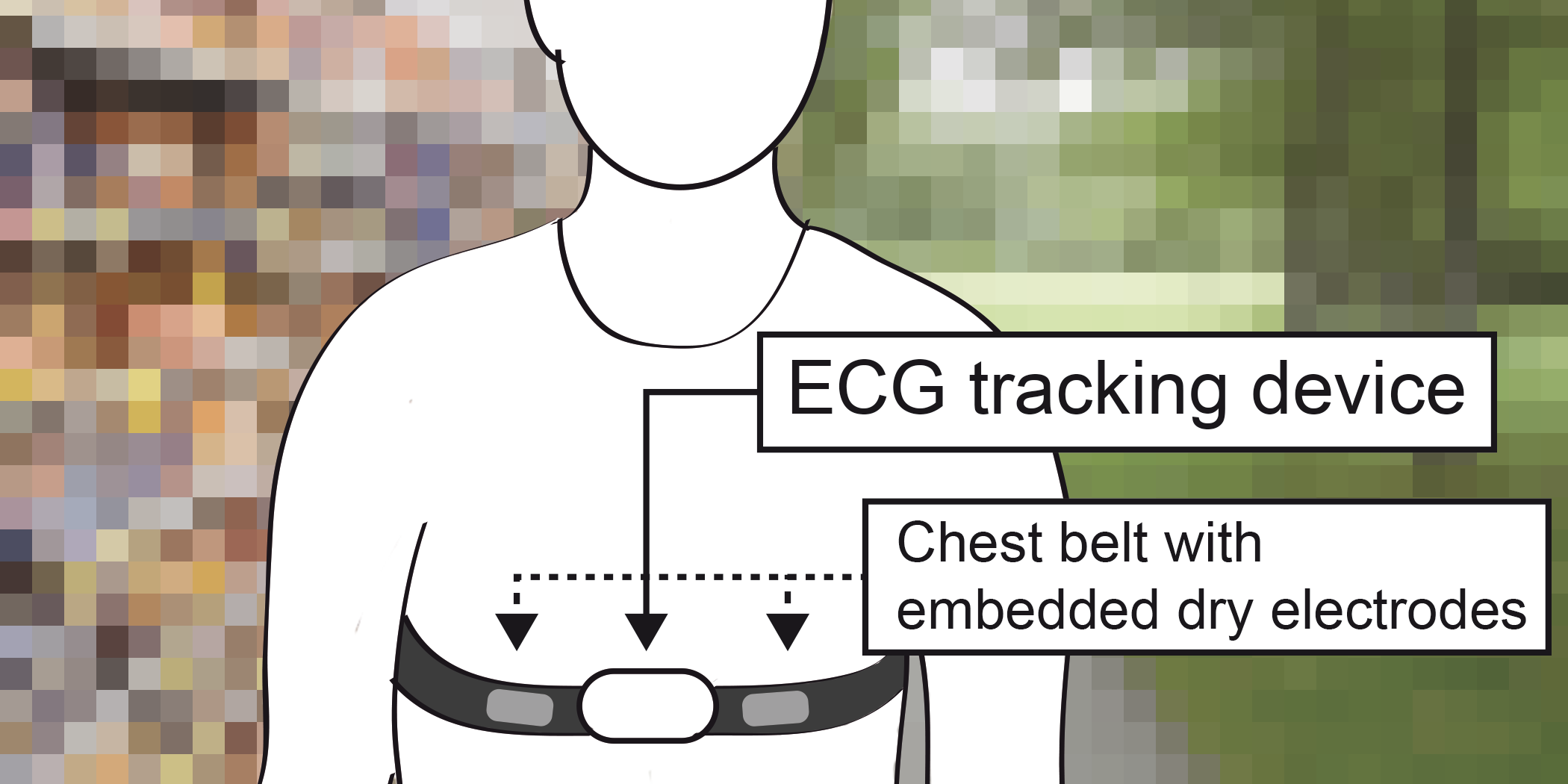}
    \caption{We conducted a field study to explore a continuous physiological signal, namely the electrocardiogram (ECG), as a biometric in everyday life. ECG data was captured from 20 people over a week, using an ECG tracker attached to a chest belt with embedded dry electrodes.}
  \label{fig:ecg-tracker-teaser}
\end{figure}

Further biometrics recently emerge in research and early applications, including electrocardiograms (ECG). An ECG is the electrical measurement of heart activity and reflects in particular the muscle contractions of the heart. This is assumed to depend on many personal factors, for example, age, gender, fitness, and genetics. While general heart functions are the same for healthy humans, considerable variation occurs due to such specific characteristics. This variation can be used to automatically identify a person, as demonstrated in first works by Biel et al.~\cite{Biel2001} and Kyoso \& Uchiyama~\cite{Kyoso2001}. Since then, ECG as a human identifier has been intensively researched in the field of biometrics (e.g. see survey~\cite{Odinaka2012}). 

However, ECG biometrics still lack in-depth evaluation in real life situations. This is crucial to provide a realistic picture of what to expect from such systems in industry and market applications. Existing work ran either controlled lab studies with medical grade ECG trackers or relied on data from medical ECG databases from PhysioNet~\cite{Goldberger2012}. In medical setups, multiple leads are used to measure the signal with adhesive electrodes attached to the body. This achieves a signal with little noise and almost no motion artifacts. In some cases the data is also manually annotated. These aspects are to be considered unrealistic and cumbersome for practical everyday applications. Some studies thus used cheaper non-medical grade tracking devices, for example, acquiring data from the fingers \cite{Arteaga-falconi2015, Chen2017, Choi2016a, Hsiao2016, Kang2016, Lourenco2011b, Pouryayevali2014}. However, only two studies went beyond the lab with wearable devices~\cite{Sprager2017, Ye2011}, with only four and five participants. We thus see the need for research to evaluate ECG biometrics in more detail ``in the wild''.

To address this gap, we conducted a field study for seven days, complemented by two lab sessions (N=20): Baseline ECG was recorded in the lab. Then, ECG was recorded in the field by equipping people with a chest belt ECG tracker worn throughout the day. We applied digital signal processing and machine learning classifiers to investigate how well user identification via ECG biometrics works on data collected in everyday life with realistic contexts.

We found that ECG biometrics can work in the field but perform less well than in controlled lab studies. We propose to combine decisions on up to 15 heartbeats to increase identification performance, achieving a best equal error rate (EER) of 9.15\%. We further show that training the model on data from a single day results in poor EER, indicating that realistic use cases should obtain training data from multiple days.

Overall, we contribute: 1) An in-depth study of ECG biometrics in everyday life, going beyond the lab. 2) Detailed evaluations of user identification on this data. In addition, we conclude with implications for future research and concrete recommendations for real world ECG biometrics systems, including critical reflections on privacy aspects. We see the need to evaluate and discuss biometric approaches not only in purely technical domains and terms but also in the HCI community, to respect user-centered aspects and concerns.

\section{Related Work}

Pioneering work in 2001 by Biel et al.~\cite{Biel2001}, Irvine et al.~\cite{Irvine2001}, and Kyoso and Uchiyama~\cite{Kyoso2001} first showed that ECG can be used to automatically identify a human individual. Their work was motivated by the fundamental question of whether ECG can be used as a biometric~\cite{Biel2001}, and more specifically whether it can secure patient data in eHealth systems~\cite{Kyoso2001}. These early studies had limited sample sizes with five to 20 participants. 

The following decade saw increasing research interest in ECG for biometrics. The survey by Odinaka et al. \cite{Odinaka2012} provides an overview and a comparison of classification approaches for user identification. Most of these projects share similar processing pipelines: First, data is collected from a study with sensors or retrieved from an existing medical database. Preprocessing then includes, for example, filtering and segmenting the signal into single cardiac cycles. Subsequently, features are extracted and used to evaluate a decision model, typically a machine learning classifier. 
The next paragraphs discuss related work for each of these steps in more detail to inform our choices in this paper.

\subsection{ECG Data Acquisition}
Public databases of signals from medical-grade devices can be found on PhysioNet\footnote{PhysioBank Databases: \url{https://physionet.org/physiobank/database}}, for example the MIT-BIH databases, which are widely used for research on ECG biometrics~\cite{Hsiao2016, Patro2017, Shen2002, Tan2017, Tantawi2015}. Alternatively, new data can be acquired from participants in experimental setups. This has been done with medical-grade devices~\cite{Biel2001}, custom prototypes~\cite{Chen2017, Hsiao2016, Kang2016, Lourenco2011b, Lourenco2012, Lourenco2014, Miao2015}, consumer devices~\cite{Arteaga-falconi2015, Choi2016a, Israel2004, Odinaka2012}, or wearables~\cite{Ye2011}. In some cases, researchers highlighted the resulting ``in-house databases'' with regard to the need for standardised datasets and evaluations for ECG biometrics~\cite{Akhter2016, Odinaka2012, Pouryayevali2014}. 

\subsection{Sensing}
Employed sensors differ in amount of leads, resolution, and sample rate: Medical-grade devices can measure ECG with twelve leads at multiple points on the body. This measures the signal from different angles facilitating diagnosis of health issues. For biometrics, one-lead measurements between the left and the right arm (i.e. body halves) are sufficient~\cite{Lourenco2011b}. Regarding sample rate, work by Seepers et al.~\cite{Seepers2017} found 90~Hz to be sufficient for authentication based on RR intervals to protect mHealth systems. They state that other work may have been oversampling the ECG by at least a factor of four. 

\subsection{Signal Processing}
Raw signals are filtered with bandpass filters~\cite{Israel2004, Israel2005, Lourenco2012, Pouryayevali2014, Tan2017, Tantawi2015}, cascaded filters~\cite{Chen2017, Choi2016a, Hsiao2016, Kyoso2001, Odinaka2010}, or finite impulse response filters~\cite{Lourenco2011b, Patro2017}. Those are mainly used to eliminate baseline drift and high frequency interference. For fiducial point extraction (i.e. detecting onset, offset, and peaks of cardiac cycles), a prior segmentation into cardiac cycles may be appropriate. Other methods like wavelet analysis do not need fiducial points at all~\cite{Tantawi2015, Ye2011}.
Segmentation typically utilises established algorithms, such as Engelse Zeelenberg~\cite{Engelse1979}, Pan Tompkins~\cite{Pan1985}, Christov~\cite{Christov2004}, or Hamilton~\cite{Hamilton2002}. 

\subsection{Feature Extraction and Classification}
Features can be extracted based on the segmented cardiac cycles: The fiducial points can be used to determine temporal features, amplitude features, as well as angles~\cite{Tan2017}. Feature representations are then utilised for decision-making via classification, including random forest classifiers (RFC)~\cite{Tan2017}, support vector machines (SVM)~\cite{Choi2016a, Lourenco2012, Lourenco2014, Ye2011}, k-nearest neighbors (kNN)~\cite{Akhter2016, Lourenco2012, Lourenco2011b}, and neural networks (NN)~\cite{Chen2017, Patro2017, Shen2002, Tantawi2015, Zhang2016}. Work by Choi et al. provides an overview, evaluating nine different classifiers for ECG biometrics~\cite{Choi2016a}.

\subsection{Example Setups and Systems}
Noteable specific classification setups include the following:
Tan and Perkowski~\cite{Tan2017} proposed to combine two classifiers, namely random forests and a wavelet distance measure with probabilistic thresholds. This improved effectiveness and robustness of their user verification system. They used data from 184 subjects under different health conditions from PhysioNet and data acquired from a biosensor integrated into a mobile device. They reported user verification accuracy of 99.52\%, which is slightly superior than each of the classifiers alone. Their conclusion points to a multimodal biometric system by complementing their approach with fingerprint biometrics.

Such a multimodal system was proposed by Hsiao et al.~\cite{Hsiao2016}, who built a prototype to collect ECG and fingerprint data. They evaluated their system on data from PhysioNet and found a recognition rate of 92\%. 

Other work addressed further specific applications such as emergency services: Ye et al.~\cite{Ye2011} acquired data from a smart textile that is capable to continuously monitor ECG. They collected around 400 hours of ECG data at 200 Hz from five firefighters in multiple sessions spread out across several months. Overall, they report a near 100\% recognition rate yet also state limitations considering the small amount of participants.

Choi et al.~\cite{Choi2016a} investigated biometric authentication with a CardioChip handheld device. They designed a cascading bandpass filter to address the higher noise of this mobile ECG sensor. Despite the mobile sensor, they collected data from 175 participants in seated position at rest for 60 seconds, with up to three repetitions. Comparing nine classifiers, they found a SVM with radial basis function kernel to perform best with an accuracy of 95.99\% and an equal error rate of 4.46\%.

Recent work also demonstrates efforts on hardware integration: Yin et al.~\cite{Yin2017} proposed a dedicated low-power ECG micro processor for ECG biometric authentication. It embeds filtering, R peak detection, outlier removal, normalisation, and authentication via neural networks. This could enable efficient biometric systems. However, it might make changes or updates to the deployed decision algorithms challenging.

\subsection{User Verification Performance and Attacks}
Overall, ECG biometrics perform well, with researchers reporting high user recognition rates, for example an accuracy of 95.99\% and EER of 4.46\% \cite{Choi2016a}, accuracy of 97.9\% and false rejection rate and false acceptance rate of almost 0\% \cite{Zhang2016}, or a recognition rate of 94.3\% and EER of 13.0\% \cite{Lourenco2011b}. When concentrating solely on HRV, research has found a recognition rate of up to 82.22\% \cite{Akhter2016}.

However, research has also shown that recognition can still be improved, for example by using subsequent heart beats~\cite{Choi2016a, Odinaka2010}: Choi et al.~\cite{Choi2016a} report a drop in EER from 4.46\% to 1.87\% when using 15 seconds of heartbeats instead of a single beat. Furthermore, other studies combined ECG biometrics with additional modalities, for example fingerprint \cite{Hsiao2016}, face recognition \cite{Israel2004}, or both \cite{Singh2012a}. Isreal et al.~\cite{Israel2004} found a multimodal biometric systems to be superior to a unimodal one.

Eberz et al. \cite{Eberz2017} demonstrated successful attacks on biometric systems that rely exclusively on ECG: They spoofed ECG signals with an attack success rate of 62\% when mapping the ECG signal from any tracking device to the target device. Huang et al.~\cite{Huang2018} relate this security issue with ECG signals to the uniqueness and stability of the features. They note that if ECG data was leaked or stolen, ECG authentication might no longer protect a system from unauthorised access. This is true for other biometrics as well, such as fingerprints \cite{Schuckers2002}.

\subsection{Methodology}
In terms of study design, a key difference between ECG biometrics studies is the number of sessions of data recording: Some studies collected data from one session only~\cite{Israel2004, Lourenco2011b, Lourenco2012, Miao2015, Patro2017}, where others used two~\cite{Arteaga-falconi2015, Chen2017, Israel2005, Lourenco2014}, or more sessions \cite{Akhter2016, Biel2001, Pouryayevali2014}. These were all lab sessions; only two studies explored ECG biometrics in the wild~\cite{Sprager2017, Ye2011}. However, these studies had severely limited samples with four and five participants, respectively. 

Regarding recording contexts and conditions, signals were obtained in resting positions~\cite{Biel2001, DaSilva2013, Kyoso2001, Odinaka2010}, and under induced strain, either physiological~\cite{Sriram2009} or psychological~\cite{Irvine2003}. 

\subsection{Summary}
In summary, ECG biometrics achieved high user recognition rates in the lab. However, evaluations under less controlled conditions are still missing. A few small-scale studies motivate more detailed investigations in real life contexts, which should include more varied contexts, body postures, and external interruptions and influences. Some researchers specifically called for such explorations of ECG biometrics in the field~\cite{Singh2012a}. 

This motivates our work: To fill the gap in the literature we present an in-depth study of ECG biometrics in the wild. Informed by the reviewed related work, we employ established signal processing procedures and compare key classification approaches (RFC, SVM, NN) on data from multiple sessions, namely seven full-day recordings per participant.
The next section presents background information on ECG signals.

\section{Background: Electrocardiogram (ECG)} \label{ECG}

\begin{figure}[tbh]
  \centering
  \includegraphics[width=1\columnwidth]{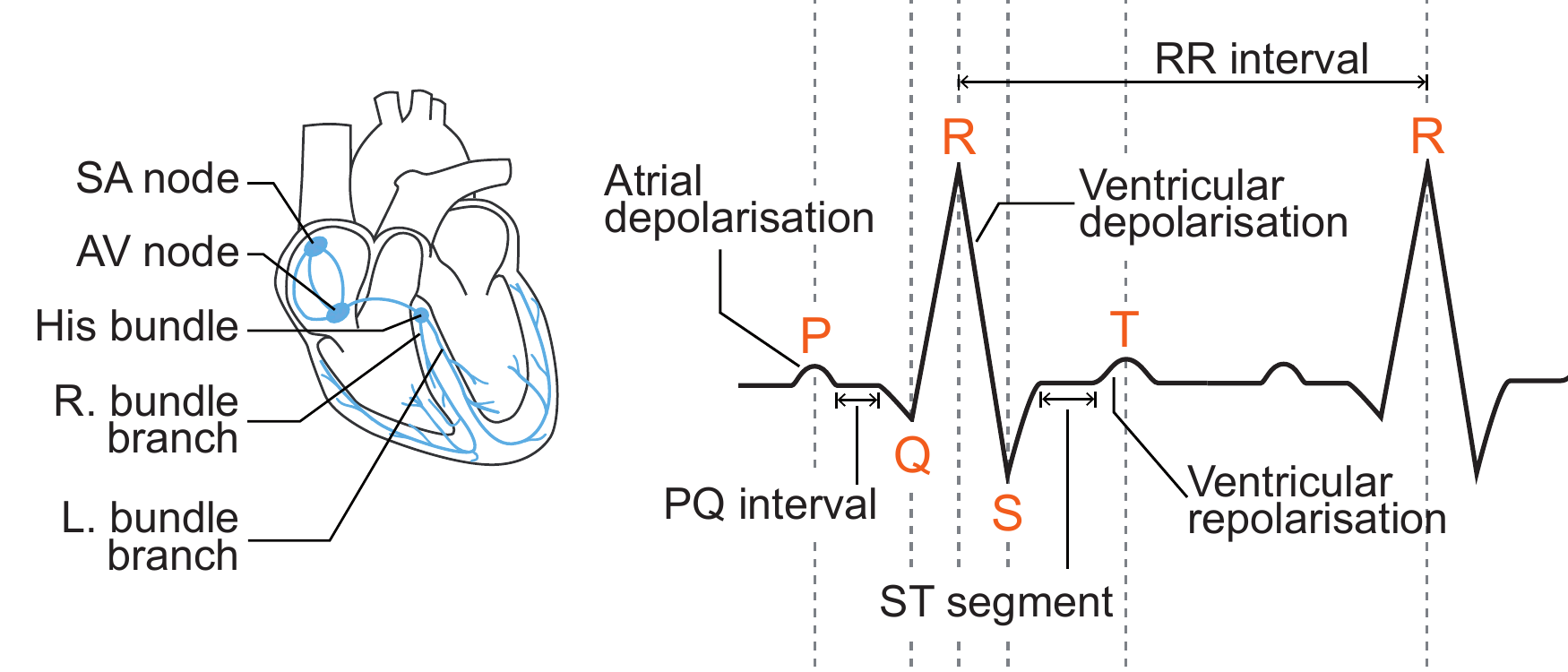}
  \caption{A schematic visualisation of the heart, with annotations of the sinus node (SA Node) and the atrioventricular node (AV Node). The SA node is the pacemaker of the heart and the AV node is important for the contractions of the ventricles. The electrocardiogram's characteristics are annotated. The P wave reflects the contraction of the atria, the QRS complex the contraction of the ventricles. The T wave signalises the normalisation of the ventricles. The RR interval is measured between two subsequent R peaks. Its variation is known as heart rate variability (HRV). Picture inspired by \protect\cite{Dubin2008, Israel2005}.}
  \label{fig:ecg-schema}
\end{figure}

The electrocardiogram reflects the electrical activity of the heart's muscle contractions. Figure~\ref{fig:ecg-schema} displays a typical cardiac cycle with its single waves annotated with letters P to T, as well as segments between some of the waves. Each wave in the ECG stands for a specific electrical activity happening in one regular heartbeat. 

\subsection{ECG Signal and Terminology}
We explain the relationship between heart activity, the resulting electrical signal, and how this is measured; with the help of work by Dubin~\cite{Dubin2008}. Details can be seen in Figure~\ref{fig:ecg-schema}:

\textit{General --} The sinus node is the natural pacemaker of the heart and triggers the cardiac cycle. In healthy humans it initiates 60 to 100 heartbeats per minute. It causes the negative charged muscle cells, so called myocytes, to change polarisation. In the resting state, the myocytes are charged negatively. 

\textit{P wave --} If the sinus node triggers a cardiac cycle, an impulse starts spreading like a circular wave and stimulates the myocytes of the atria to be positively charged (depolarisation). As the wave spreads, the atria contracts. This causes the P wave on the ECG. 

\textit{PQ interval --} The depolarisation wave reaches the atrioventricular node, where it is slowed down, showing in the signal as a short pause. The time between the P and the Q wave is needed for the blood to flow from the atria into the ventricles. After the depolarisation has passed the AV node and reaches the His bundle it shoots through the right and left bundle branches.

\textit{QRS complex --} As soon as the depolarisation shoots through bundle branches, the ventricle's myocytes change their charges. The ventricles contract and blood flows into the system. 

\textit{ST segment --} Directly after the QRS complex, the ventricles start to repolarise. Within the ST segment the ventricular repolarisation is almost not visible in the signal.

\textit{T wave --} After the ST segment, the ventricular repolarisation happens more intensively, causing the T wave. The whole cycle will again be initiated, after a short pause. Note, that atria have already repolarised, but the atrial repolarisation has a low amplitude and is overridden by the QRS complex.

\textit{RR interval --} The RR interval is measured between two successive heartbeats and determines a pulse. The variation of multiple RR intervals is known as heart rate variability (HRV). 

\subsection{Measuring ECG} 
ECG is measured with electrodes attached to the human body. In particular, the amplitude is measured as voltage over time. A simple ECG can be measured with pairs of electrodes, a lead can be interpreted as a line between two electrodes. The bipolar limb leads are measured as follows: Lead 1 is measured between both arms, Lead 2 is measured between the right arm and the left leg, Lead 3 is measured between the left arm and the left leg. Unipolar leads are measured between both arms and the left leg. However, electrodes and cables are combined differently than in the bipolar setup. In medical setups up to 12 leads are used to record an ECG. These measurements include the bipolar limb leads, unipolar (augmented) limb leads and six chest leads. This way the ECG can view the cardiac cycle from 12 different angles. This can help to diagnose serious health conditions. In constrast, non-medical ECG sensors mostly measure the ECG only with one pair of electrodes horizontally between the body halves (Lead 1).

\section{Field Study: ECG in the Wild}

Here we describe our field study. We aim to answer questions on two scenarios: 1) From lab to the wild: How well do ECG biometrics work on data recorded in everyday life in general? 2) Usable setup: How well do ECG biometrics work in the wild with a setup that respects basic usability considerations for everyday applications?

\subsection{Participants}

We recruited 21 participants (9 female, 12 male) via university newsletters and social media. One male participant dropped out due to a serious health condition while the study was running (unrelated to study participation). Thus, 20 participants (11 male, 9 female) completed the study. Their age ranged from 18-34 years, with a median of 27 years.

Self reports of height and weight were used to calculate the body mass index (BMI) which ranged between 19.46 and 38.81 kg/m$^2$. According to the BMI, two participants would be considered as overweight and two as obese. No one reported a diagnosed heart condition.

Participants provided basic information on their lifestyle: Two reported to not do any sports, one to exercise twice a month, while the remaining 17 exercised once a week or more (e.g. going for a run for 30 minutes). Regarding the use of substances which might influence the cardiovascular system, six participants reported to be regular smokers, 17 to occasionally drink alcohol, and five to irregularly use medication.

People were compensated with a 30 EUR gift card or cash.

\subsection{Apparatus}

The ECG signal was recorded with a chest belt tracking device (EcgMove3 \& EcgMove4\footnote{movisens: \url{https://www.movisens.com/}}). We had one EcgMove4 and eight EcgMove3. Both models provided all relevant features for our study and the manufacturer ensured that both versions use the same ECG sensor electronics. ECG is recorded with a resolution of 12 bit and a sample rate of 1024 Hz. A battery charge lasted for about two days.

According to the manufacturer, the tracker must be worn with a belt around the chest as shown in Figure~\ref{fig:ecg-tracker-teaser}. The belt should fit tight to the body, in such a way that the embedded dry electrodes are in contact with the skin. Body hair may be suboptimal for wearing the tracker. No participant reported that this was an issue.

\begin{figure}[t]
  \centering
  \includegraphics[width=1\columnwidth]{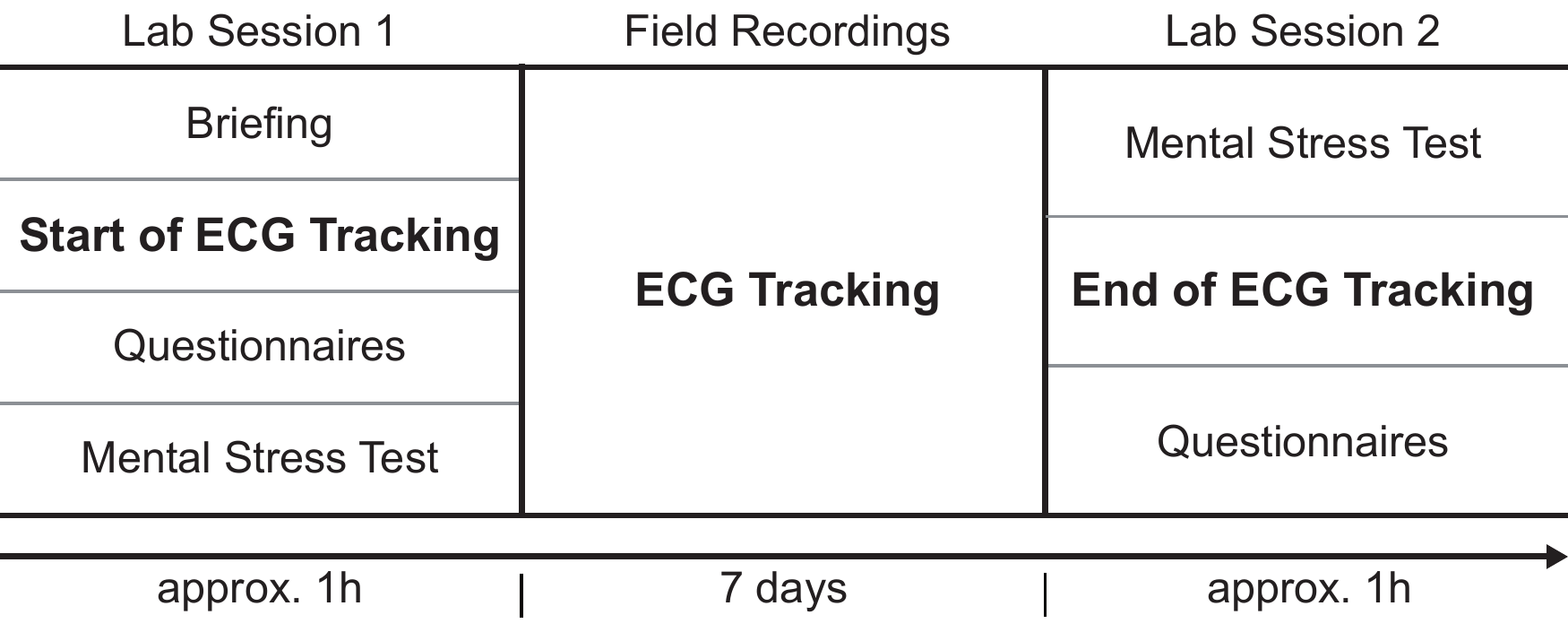}
  \caption{Study procedure: Read left to right, lab sessions top to bottom. As we focus on the analysis of ECG as a biometric in this paper, ECG is bold in the figure. Mental stress tests were carried out to study the influence of mental stress on the ECG, but are not part of this paper.}
  \label{fig:procedure-complete}
\end{figure}

\subsection{Procedure}

The study procedure is depicted in Figure~\ref{fig:procedure-complete}: Each participant attended two lab sessions, one at the start and one at the end of the recording phase. One lab session took approximately one hour and included a mental stress test which we do not analyse in this paper. Inbetween the lab sessions, ECG was recorded in the field. This way, we recorded six complete days, and two fractions of a day (start and end day) for each participant. We consider recordings in the lab as an ECG baseline for biometrics, as the measures took place under controlled conditions. 

Participants were instructed to wear the ECG tracker when awake. Exceptions were highly intensive sports and showering or taking a bath. Regarding the manufacturer, the sensors can be worn when doing sports. However, we decided to instruct participants not to do so since we judged it as uncomfortable based on a pretest, also considering transpiration. 

\section{Data Processing}

We collected about 36 GB of raw tracking data. From this data we consider in particular the ECG, the charging state, and an internal measure of the tracker that reflects the validity of the ECG. We describe our preprocessing steps as follows.

\subsection{Step 1: Converting Files}

Initially, the data was extracted from the trackers with the manufacturer's tool SensorManager\footnote{SensorManager: \url{https://www.movisens.com/en/sensormanager}}. The resulting files were then converted into the European Data Format (EDF+)\footnote{EDF+: \url{https://www.edfplus.info}}, which stores multichannel biological and physical signals. For example, it is commonly used for ECG, EMG, or EEG data. This format allowed us to seamlessly process the data with existing standard libraries and tools. Conversion to EDF+ almost tripled the file size.

\subsection{Step 2: Downsampling}

To enable efficient data analysis, we downsampled the data from 1024 Hz to 256 Hz using the tool EDFBrowser\footnote{EDFbrowser: \url{https://www.teuniz.net/edfbrowser/}}, which automatically anti-aliased the signals as well. A sample rate of 256 Hz should be sufficient for ECG biometrics according to the literature \cite{Israel2004, Pouryayevali2014, Tantawi2015, Ye2011}. The effects of downsampling are illustrated on an example heartbeat in Figure~\ref{fig:preprocessing-signal} (left).

\subsection{Step 3: Removing Charging States and Invalid Measures}

The single recordings were still too large to process on our machine with 16 GB RAM. Therefore, we split the complete recordings into smaller chunks.

We also removed parts of each day where the device was either charging or did not detect a valid heart rate according to its internal validity check. At these points the files were split into smaller files to store clean episodes directly.

\subsection{Step 4: Filtering}

Technically, filtering and segmentation (see next step) were carried out in one go. For readability we report them separately here. Both were executed using the neurokit library for Python\footnote{NeuroKit: \url{https://neurokit.readthedocs.io}}, which combines several libraries such as biospy\footnote{BioSPPy: \url{https://biosppy.readthedocs.io}} for processing physiological signals like EEG or ECG. 

\begin{figure}[bt]
  \centering
  \includegraphics[width=1\columnwidth]{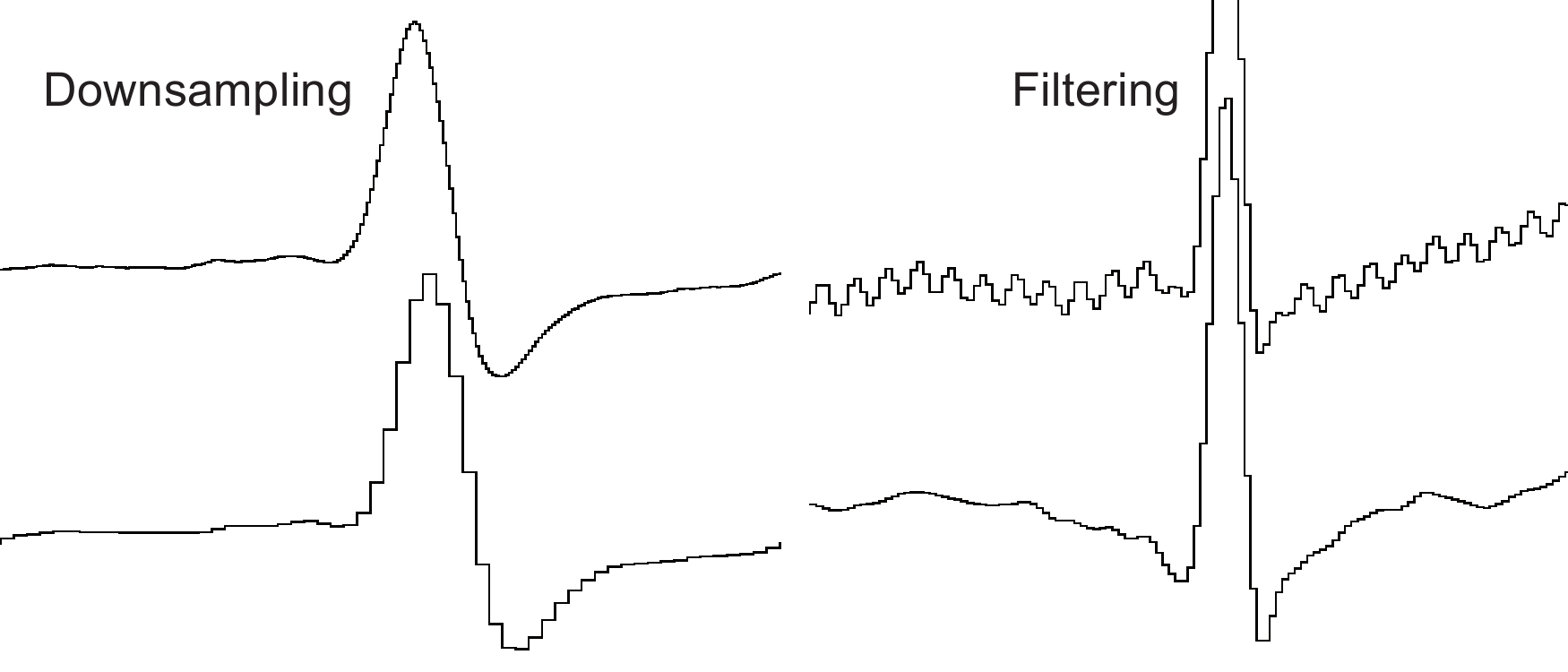}
  \caption{Signal preprocessing. Left: An example of downsampled data. The top signal has 1024 Hz, the bottom one 256 Hz. Right: Examples before and after the filtering. The top plot shows an unfiltered signal with visible interferences. The plot below shows the filtered signal. 
  }
  \label{fig:preprocessing-signal}
\end{figure}

Visual inspection of the ECG showed noisy episodes in the recordings. For example, we found high frequency interferences on some parts of the signals and minimal baseline drift. We thus applied a finite impulse response (FIR) bandpass-filter, configured with 77 taps and cutoff frequencies of 3 and 45Hz. Only parts of the signals within that range pass the filter. For example, any high frequency larger than 45Hz will be eliminated. An example displaying the effects of the filter can be seen in Figure~\ref{fig:preprocessing-signal} (right).

\subsection{Step 5: Segmentation (Peak Detection)}

To segment the recordings into single cardiac cycles, we applied R peak detection as proposed by Engelse and Zeelenberg~\cite{Engelse1979}, with modifications by Lourceno et al.~\cite{Lourenco2012}. Cardiac cycles and the other peaks -- such as P, Q, S, and T -- were then extracted based on the R peaks. In particular, neurokit uses fixed thresholds to determine cardiac cycles and to search for maxima within the cycle to detect the other peaks.

\subsection{Step 6: Outlier Dectection \& Removal}

The detected cycles and peaks were checked for outliers. For this we first computed intervals between the peaks: PQ, QR, RS, QS, and ST. A cardiac cycle was considered an outlier when either 1) not all intervals could be calculated due to the lack of the correct detection of peaks, or 2) an interval was further than three standard deviations away from the mean interval (specific to each type of interval). The files were split at points where an outlier was found to store clean episodes.

\subsection{Step 7: Feature Computation}

Checking random samples of the previously determined intervals showed high deviations for the PQ and the ST intervals. Additionally, visual inspection of the detected peaks showed that the algorithms had problems with correctly identifying the P and T peaks. Specifically, this happened when the amplitudes of the waves were rather low and could not be differentiated from surrounding noise. For example, see the bottom right signal in Figure~\ref{fig:preprocessing-signal}. 

Due to these difficulties with the PQ and ST intervals we focused on the QRS complex and derived statistical measures based on three subsequent samples (heartbeats): min, max, mean, median, standard deviations. Those measures were computed for the QR, RS, and QS intervals. Intervals were computed peak to peak. Selecting the best features for identification was part of the evaluation, see section \nameref{label:feature_selection}.

\subsection{Step 8: Composing a Dataset for Evaluation}
The number of recorded datapoints varied between participants. Thus, we subsampled the data per participant to create a dataset with a roughly comparable number of samples for each person. This helps with our evaluation schemes later on and facilitates comparisons with related work and dataset sizes reported therein. We ensured to evenly include data from each person and day.

In this way, we sampled a total of 50,000 datapoints per participant across the entire period of the study.

\section{Evaluation Setup}
We next describe our evaluation goals and procedures.
We consider two questions and corresponding scenarios:

\textit{Scenario 1 -- from lab to the wild: How well do ECG biometrics work on data recorded in everyday life in general?} With this scenario we test if ECG biometrics generally work on everyday life data, recorded outside of the controlled lab setups of previous work. Here we train the system on data from all days of the study, which means that the system has also seen data from the same days as it is then tested on. While somewhat unrealistic, this facilitates fundamental comparisons to related work which recorded data in a single sitting.

\textit{Scenario 2 -- usable setup: How well do ECG biometrics work in the wild with a setup that respects basic usability considerations for everyday applications?} Post-hoc data analysis sometimes tends to miss out on the implications that some evaluation schemes have for practical deployments, as highlighted in related work~\cite{Buschek2018}. In particular, for ECG biometrics users would need to provide some training data initially (enrollment phase). With this second scenario we thus evaluate user authentication in a more realistic approach with regard to such implied enrollment needs for users: We assume that enrollment takes place on one/two/three starting day(s) only and the system then has to still work on the following days. In practice, this would minimise time and user effort required for training the system.

\subsection{Basic Cross-Validation Procedure}
In general, following typical biometrics evaluations, we test if a system can distinguish between different people based on the recorded ECG data. Intuitively, we train classifiers to distinguish between one legitimate user and everyone else, and then test this system with data from that legitimate user and a person that the classifier has not seen before (``attacker'').

\begin{figure}[tb]
  \centering
  \includegraphics[width=0.85\columnwidth]{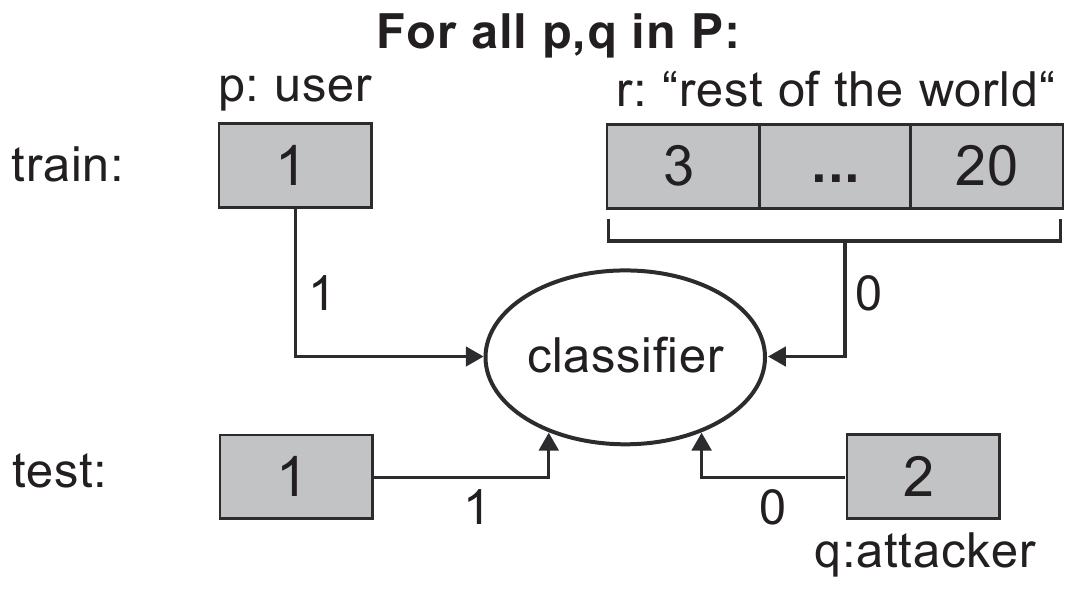}
  \caption{The evaluation scheme used to evaluate classifiers for user authentication on our ECG data, following the recommended procedure in related work~\protect\cite{Buschek2015chi}. See text for details.}
  \label{fig:model}
\end{figure}

This evaluation scheme is depicted in Figure~\ref{fig:model}. Formally, for each tuple of two participants $(p,q)$ we assign $p$ as the legitimate user and $q$ as the attacker. All other participants $r \in P \setminus \{p,q\}$ mimic the ``rest of world'', for example all other users within an enrollment database. We train a classifier $c_p$ to separate two classes: The legitimate user $p$, which is labeled as 1, and all other users $r$, which are labeled as 0. We then test this classifier by feeding it 1) the test part of $p$'s data, which it should ideally accept (output 1), and 2) the attacker data ($q$'s data), which it should ideally reject (output 0). 
This scheme follows the recommendation in related work~\cite{Buschek2015chi}, which showed that excluding the attacker from the rest-of-world group is important to avoid unrealistic and biased results.

\subsection{Training \& Test Sets}
We generated training and test sets for each person-attacker combination as described above. In both scenarios, attacker data was spared out from training. The particular choice of data points differed between the two scenarios.

\subsubsection{Scenario 1} 

Here we chose a random subsampling approach, where the order of days or cardiac cycles does not need to be considered. For each user-attacker combination we picked a fixed number of datapoints at random from the legitimate user's samples, and split those into training and test sets. For the rest of the world, we did the same but ensured that data is evenly picked from data for each day and participant. Finally, we picked a random sample of data from the attacker for testing.

For each user-attacker combination this resulted in 4,000 samples for both user and rest-of-world. Test data had 1,000 samples for both user and attacker. This gives 8,000 samples for training and 2,000 samples for testing, reflecting a typical 80 / 20 train-test split. The total training data is equivalent to 24,000 cardiac cycles, while the testing data corresponds to 6,000 cycles. If we consider an average heart rate of 60 bpm, this results in evaluations using 6.66 hours of training data and 1.66 hours of test data per user-attacker combination. 

\subsubsection{Scenario 2}

For scenario 2 we defined a more realistic use case where enrollment is assumed to happen on a starting day and the system is validated against data of the following days. Thus, keeping the order of days and cardiac cycles is crucial. For each user-attacker combination, training data for the user was taken from day two, as this was the first day where recordings happened across the whole day (see Figure~\ref{fig:procedure-complete}). Training data for the ``rest of the world'' was evenly sampled from all days. The user's test data was sampled beginning with day three, whereas the attacker's test data was evenly taken from all days. While we kept the sample order, note that samples might be joined across temporal gaps in the recordings (e.g. gaps due to taking off the tracker for doing sports; also see step 8 in the Data Processing Section). 

In the following we are going to call this scenario ``scenario 2a'' (considering enrollment data from only one day). For closer investigation of the effect of the number of days used for enrollment, we further adapted scenario 2a. We extended the training data by a second and third day, that is, data from study day three and day four. Accordingly, we removed day three, respectively day four, from the test data. These adaptions are called scenario 2b (two days for training) and scenario 2c (three days for training).

For each user-attacker combination this resulted in 2,000 samples for both user and rest-of-world, while test data had 500 samples for both user and attacker. This is less data than in scenario 1 because some participants had less than 4,000 samples on day two. Scenario 2b and 2c were not affected by this reduction. Since one participant's data was too fragmented and contained only few samples because of very noisy signals. We decided to leave out this participant from evaluation for both scenarios. 

In scenario 2a the data is equivalent to 16,000 cardiac cycles for training, and 3,000 cycles for testing. Again considering an average pulse of 60bpm this results in evaluations using 4.44 hours of training data and 0.83 hours of test data per user-attacker combination. For scenario 2b and 2c amount of data was equal to scenario 1.

\subsection{Feature Selection}
\label{label:feature_selection}
We identified the best features to use by computing a correlation matrix. Additionally, we computed feature importances with the help of an Extra Tree Classifier with 100 estimators. This way we identified the min, max, and mean of the QR, RS, and QS intervals as the most promising features.

\subsection{Performance Measures}

We evaluated three classifiers: A Random Forest Classifier (RFC), a Support Vector Machine (SVM), and a Neural Net (NN). We used implementations from the Python libraries scikit-learn\footnote{scikit-learn: \url{https://scikit-learn.org/stable/}} v0.20.2 and tensorflow\footnote{TensorFlow: \url{https://www.tensorflow.org/}} v1.12.0. 

In the following, we report the equal error rate (EER), and use the receiver operating characteristic (ROC) curve for visualisation. Both EER and the ROC curve are measures that depend on the tradeoff between true positive rate (TPR) and false positive rate (FPR). For both, the threshold applied to the model's output (i.e. class probability / score) is varied. The ROC curve visualises this by plotting the resulting pairs of TPR and FPR values. The EER describes the point where TPR and FPR are equal. We decided for the EER and ROC curve as they are commonly used in related work, provide a one number summary, and visualise the overall accuracy of classification with probabilities on a two-class problem. Thus, they can be used to interpret the quality of a biomteric system.

\subsection{Classifiers}

We used the libraries' default hyperparamter settings for the classifiers, with some adaptions. These were informed by cross-validation of a small set of values and NN architecture variations, using a small randomly selected part of the data. We utilised the sklearn library for the Random Forest Classifier with 100 estimators, and a Support Vector Machine with linear kernel, weighted classes, and C=10e-4. Tensorflow was used for the Neural Net, which consisted of three dense layers with 32 units each and ReLu activation, one binary output layer with sigmoid activation; Adam optimizer with a learning rate of 10e-4; binary crossentropy loss. 

For scenario 1, 100 training epochs turned out to be sufficient to train the NN. We plotted the loss and accuracy for training and validation. A visual inspection of the plots showed that the curves started to stabilise after epoch 50. For scenario 2 we found 25 training epochs to be sufficient. After that, training and validation loss curves started to drift apart, which typically indicates overfitting.

\begin{table*}[t!]
\centering
\small
\begin{tabular}{lcccccccc}
\toprule
\multicolumn{1}{c}{\multirow{2}{*}{Classifiers}} & \multicolumn{2}{c}{Scenario 1} & \multicolumn{2}{c}{Scenario 2a} & \multicolumn{2}{c}{Scenario 2b} & \multicolumn{2}{c}{Scenario 2c} \\
\multicolumn{1}{c}{}                             & \small{ROC AUC}       & \small{EER (\%)}       & \small{ROC AUC}                & \small{EER (\%)}               & \small{ROC AUC}                & \small{EER (\%)}                & \small{ROC AUC}                & \small{EER (\%)}                \\
\midrule
RFC                                              & 0.908         & 16.68          & 0.843                  & 21.91                  & 0.859                  & 20.80                   & 0.877                  & 19.54                   \\
NN                                               & 0.898         & 17.70          & 0.800                  & 26.17                  & 0.859                  & 21.02                   & 0.876                  & 20.16                   \\
SVM                                              & 0.805         & 25.70          & 0.759                  & 28.08                  & 0.771                  & 28.10                   & 0.784                  & 26.77             \\
\bottomrule
\end{tabular}
\caption{Summary of the classification results found in our scenarios. With scenario 1 we investigated if ECG biometrics generally work in the field. Training and test data was taken from all days. In contrast, in scenario 2a training data was taken from the first complete day, test data from the remaining days. This reflects a more realistic use case where enrollment happens on one day and authentication on subsequent days. Scenarios 2b and 2c are variations of scenario 2a: Here we varied the amount of training and test days. For scenario 2b we used data from two days for enrollment, for 2c we used data from three days for enrollment.}~\label{tab:performance_classifiers}
\end{table*}

\begin{figure*}[t!]
\vspace{-2.25em}
\centering
    \subfloat[]{\includegraphics[width=0.25\textwidth]{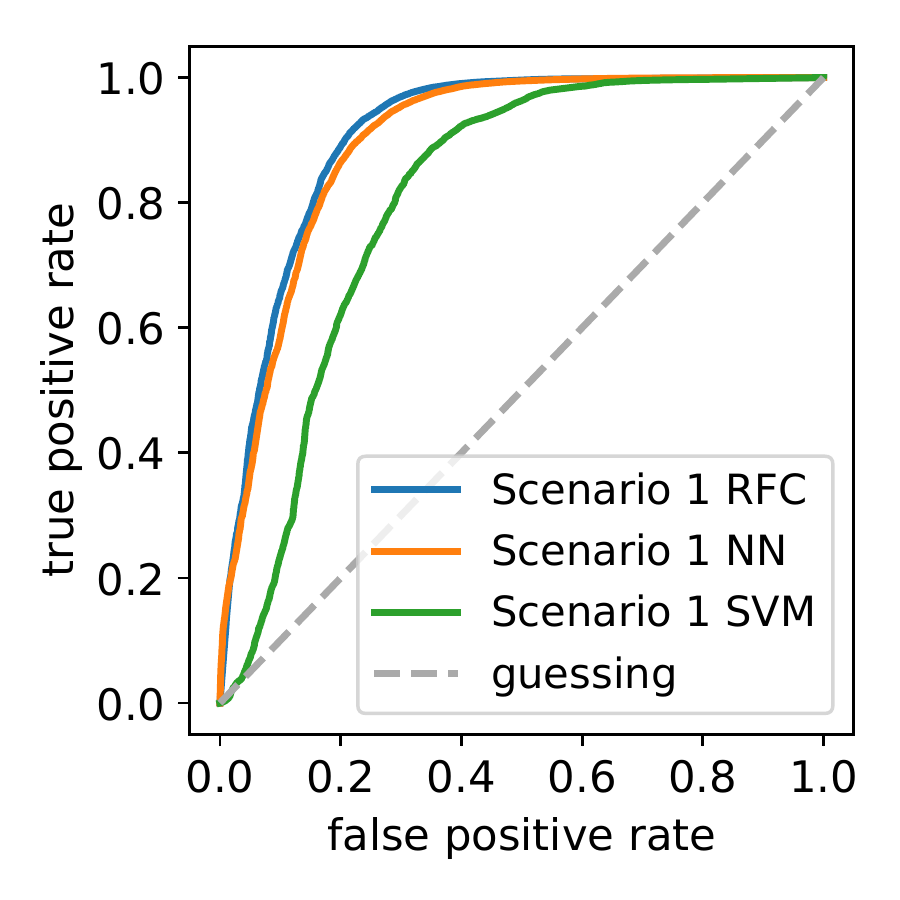}}
    \subfloat[]{\includegraphics[width=0.25\textwidth]{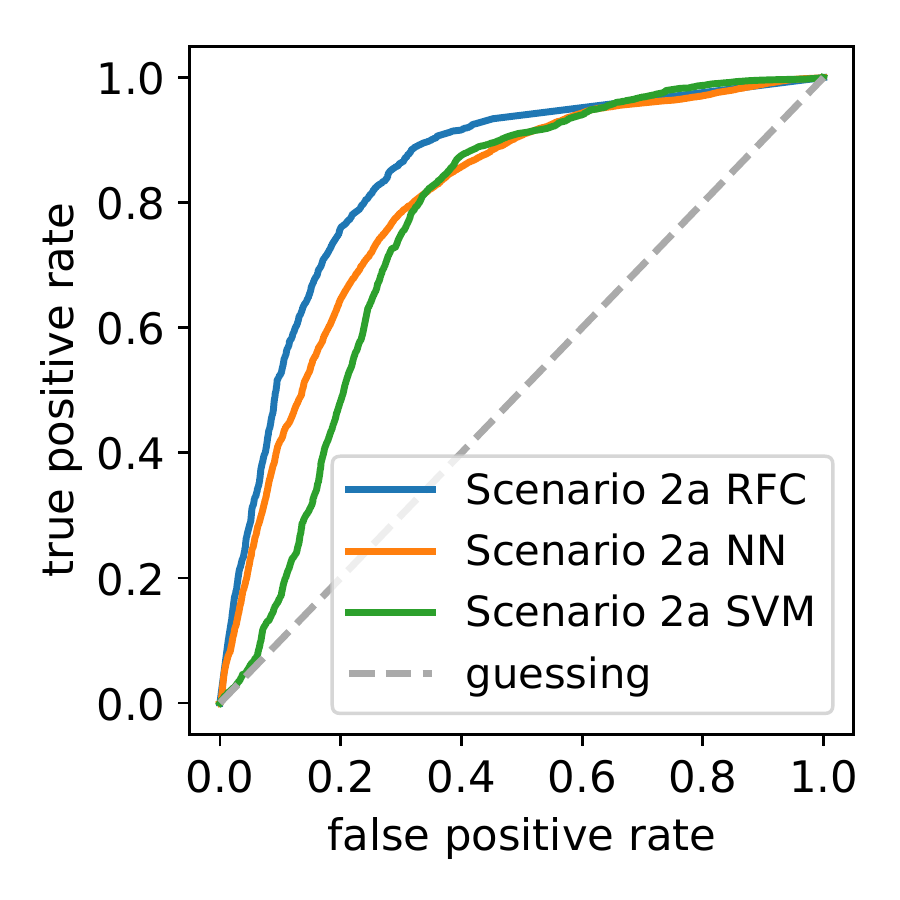}}
    \subfloat[]{\includegraphics[width=0.25\textwidth]{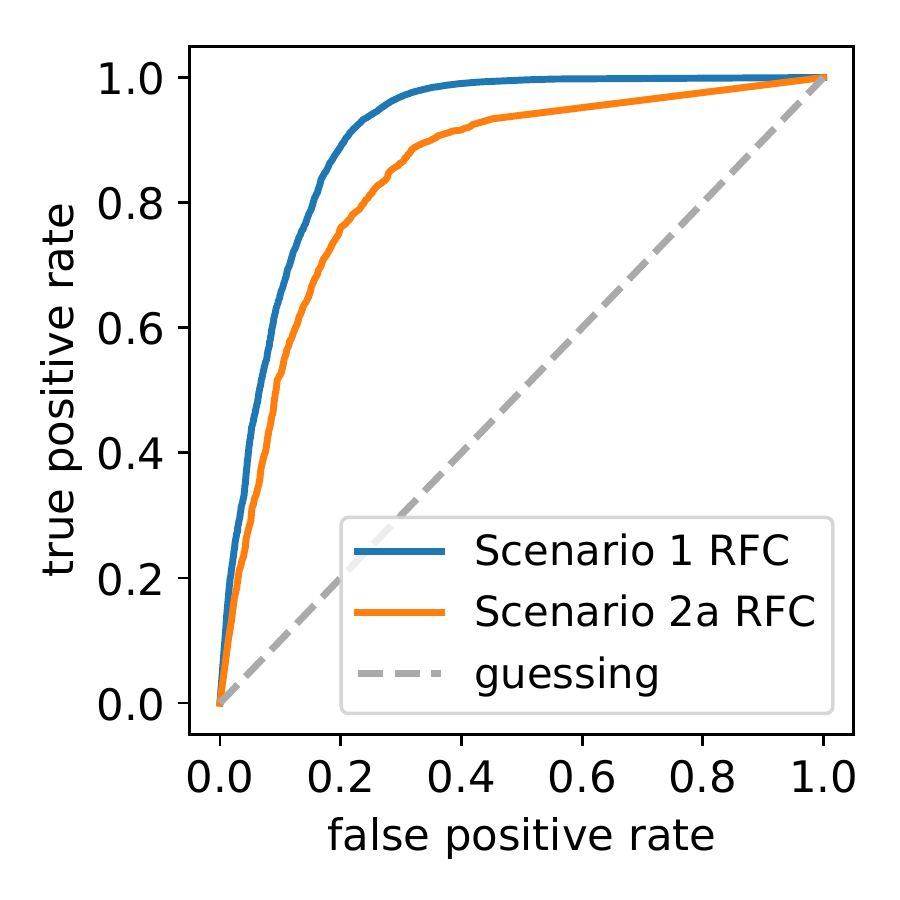}}
    \subfloat[]{\includegraphics[width=0.25\textwidth]{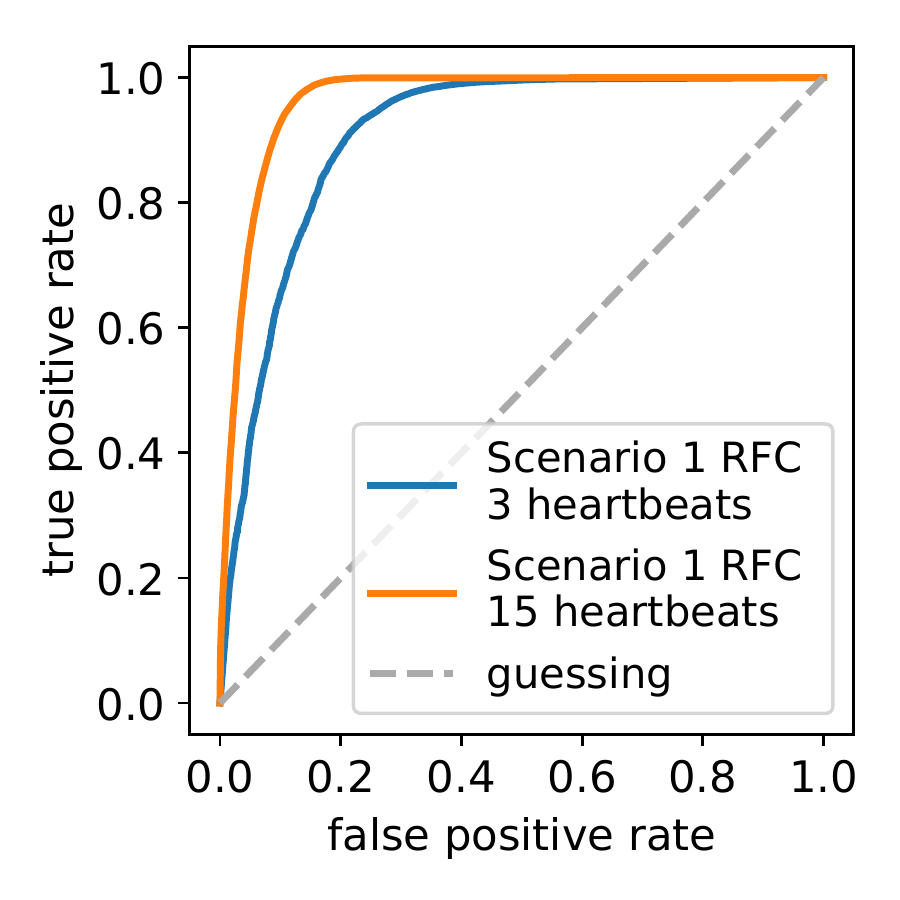}}
   \caption{Receiver operating characteristic (ROC) curves, visualising a) Comparison of classifiers in scenario 1; b) Comparisons of classifiers in scenario 2a; c) Comparison between scenario 1 \& scenario 2a for the random forest classifier; d) Comparison of decisions made by the random forest classifier on 3 and 15 heartbeats in scenario 1. Values for the ROC area under curve (AUC) and the equal error rate (EER) for a), b), and c) can be seen in Table~\ref{tab:performance_classifiers}. Values for d) can be seen in Table~\ref{tab:combined_decisions}.}
  \label{fig:rocs}
\end{figure*}

\section{Results}

\subsection{Basic Results}

For scenario 1 we found an ROC AUC of 0.908 and an EER of 16.68\%, based on the decisions of the RFC. The NN performed almost comparable with an ROC AUC of 0.898, and an EER of 17.70\%. The SVM was worst with an ROC AUC of 0.805 and an EER of 25.70\%. Results are summarised in Table~\ref{tab:performance_classifiers} (see column scenario 1). For a visualisation of the ROC curves in scenario 1 see Figure~\ref{fig:rocs}a.

For scenario 2a we found a ROC AUC of 0.843 and an EER of 21.91\% for the decisions made by the RFC. The NN is inferior to this with a ROC AUC of 0.800 and an EER of 26.17\%. Again, the SVM yields the worst performance with a ROC AUC of 0.759 and an EER of 28.08\%. A comparison can be seen in Table~\ref{tab:performance_classifiers} which summarises the results (see column scenario 2a). The corresponding ROC curves for scenario 2a are depicted in Figure~\ref{fig:rocs}b.

\subsection{Comparison of Scenarios \& Influence of Enrollment Days}

Using one day for enrollment in scenario 2a shows worse performance than scenario 1. For example, comparing the RFC between the scenarios we found an absolute difference of 0.065 for ROC AUC and 5.23\% for ERR, see Figure~\ref{fig:rocs}c. Using three days for enrollment in scenario 2c we found an ROC AUC of 0.877 and an EER of 19.54\%. The difference between scenario 1 and scenario 2c then was 0.031 for ROC AUC and 2.86\% for EER. This improvement might be due to variance in heartbeats and everyday contexts which can not be captured within a single day.

\subsection{Combining Multiple Decisions}

The previous results were obtained on classification of single datapoints, that is, feature vectors derived from three subsequent heartbeats. We also evaluated making decisions based on a combination of subsequent multiple such datapoints. In line with related work~\cite{Choi2016a}, we combined up to five subsequent decisions. This is equal to 15 heartbeats, or 15 seconds, assuming a normal heartrate of 60bpm. Using decisions by the RFC in scenario 1, we found this to improve absolute ROC AUC by 5.6\% and absolute EER by 7.53\%, as seen in in Table~\ref{tab:combined_decisions} and Figure~\ref{fig:rocs}d. Applying the same approach for scenario 2a, we found an improvement in absolute ROC AUC of 5.5\% and absolute EER of 5.29\%. Decisions using three heartbeats in scenario 1 perform nearly as well as decisions using 15 heartbeats in scenario 2a, which can be seen in Table~\ref{tab:combined_decisions} as well.

\begin{table}[tbh]
\centering
\small
\begin{tabularx}{\columnwidth}{lcccc}
\toprule
\multirow{2}{*}{Heartbeats} & \multicolumn{2}{c}{Scenario 1 (RFC)} & \multicolumn{2}{c}{Scenario 2a (RFC)} \\
                            & \small{ROC AUC}          & \small{EER (\%)}          & \small{ROC AUC}          & \small{EER (\%)}          \\
\midrule
3                           & 0.908            & 16.68             & 0.843            & 21.91             \\
6                           & 0.939            & 12.76             & 0.874            & 18.76             \\
9                           & 0.952            & 10.98             & 0.886            & 17.96             \\
12                          & 0.959            & 9.83              & 0.894            & 16.96             \\
15                          & 0.964            & 9.15              & 0.898            & 16.62             \\
\midrule
Diff.                 & 0.056            & 7.53              & 0.055            & 5.29          \\
\bottomrule
\end{tabularx}
\caption{Results of combined decisions made on subsequent heartbeats by the random forest classifier (RFC) in scenario 1. In total the EER decreased by 7.53\% in scenario 1 (i.e. from 3 to 15 heartbeats). In scenario 2a the EER decreased in total by 5.29\%.}
\label{tab:combined_decisions}
\end{table}

\subsection{Subjective Feedback}
\begin{figure}[tbh]
  \centering
  \includegraphics[width=1\columnwidth]{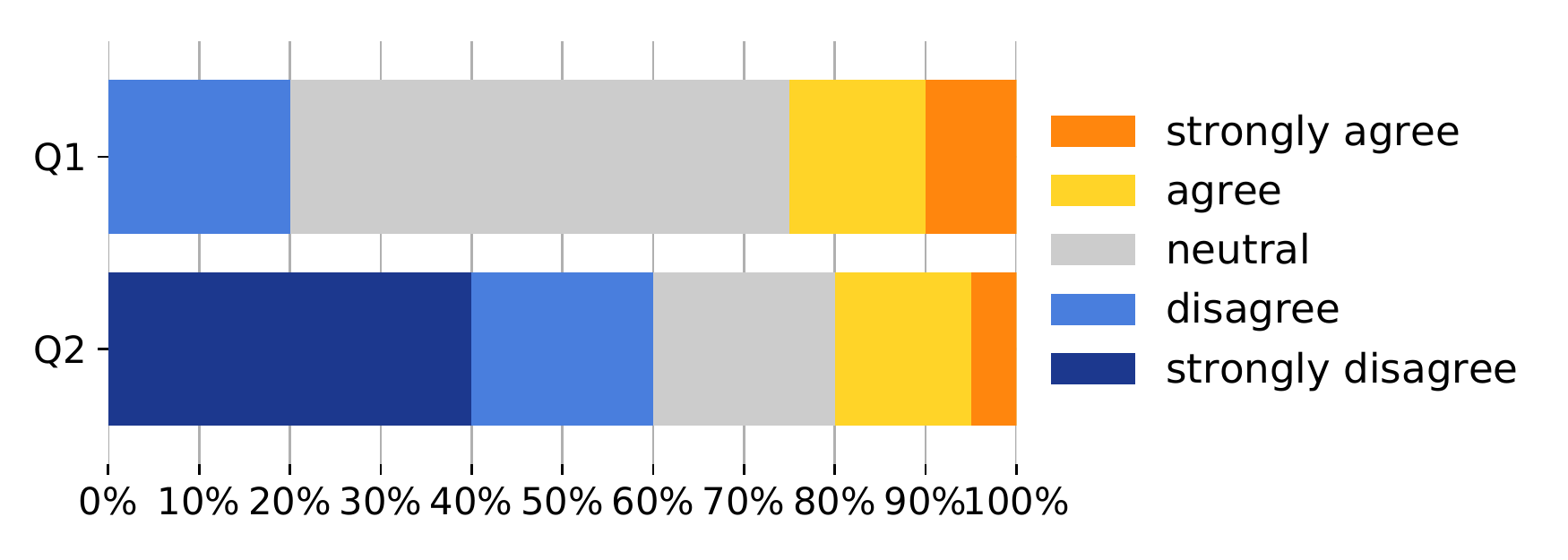}
  \caption{Results of the subjective feedback regarding ECG biometrics in everyday life. Q1: It is comfortable to wear the chest belt tracking device. Q2: I would like to wear an ECG tracker for biometric purposes.}
  \label{fig:subjective-feedback}
\end{figure}

We asked participants for subjective feedback at the end of the study. This included two five-point Likert questions considering ECG biometrics. In particular, we asked people to rate their agreement to the comfort of wearing the chest belt and how much they would like to wear an ECG tracker for biometric purpose. Figure~\ref{fig:subjective-feedback} shows the results.

The comfort of the chest belt was rated as neutral by more than half (55\%) of the participants; 20\% found it uncomfortable, 15\% comfortable and 10\% very comfortable. No one rated it very uncomfortable.

The desire to wear an ECG tracker for biometric applications was rather small: 40\% rated it very unlikely, 20\% unlikely, 20\% neutral, 15\% likely, and 5\% very likely. 

\section{Discussion}

We discuss our findings with regard to related work, implications for future research and real world applications, as well as critical reflections on privacy in this context.

\subsection{Distinguishing People based on Real World ECG Data}

Our results show that it is generally possible to distinguish people by their heartbeats recorded with a non-medical ECG tracker in everyday life. In particular, we found a best EER of 9.15\% when authenticating users based on 15 heartbeats and including training data from each day (scenario 1). 

For three heartbeats, we found a higher EER of 16.68\%. This is higher than typical numbers in related work. We explain this with the challenges of varying contexts and higher noise of real world recordings, compared to controlled lab settings. Thus, we conclude that future research should evaluate ECG biometrics systems in their intended application contexts, not only in lab settings, to avoid overly optimistic results.

Regarding uniqueness of the ECG, we found no significant difference in EER with altering group sizes (5, 10, 15), which was also shown by Carreiras et al. \cite{Carreiras2016}. Moreover, a first comparison in our data showed 5.85\% of pairs of participants to be hard to be distinguished from each other. The threshold for this was the mean EER of all pairs plus two times the standard deviation. For those ``difficult pairs'' we could suspect some similarities in physiology and lifestyle yet investigating this would require more data (e.g. magnetic resonance imaging, experience sampling). We thus suggest to further explore uniqueness of ECG in the future, also with larger samples.

\subsection{Comparison to Related Studies}
In the following, we discuss our results in the light of key related work. Note that direct comparisons of one-number summaries such as EERs are difficult due to study differences. Thus, we do not interpret \textit{exact} values but rather discuss the wider picture.

A number of papers report lower EERs in lab studies: For example, work by Choi et al.~\cite{Choi2016a} found an EER of 4.46\%. They recorded 60 seconds of ECG data per participant in the lab, in a seated position. They trained their classifiers on the first 30 seconds, while the rest was used for testing. In practical biometric security applications, enrollment and authentication are not likely to happen within a single minute. Together with the more controlled setup, this likely explains the better performance compared to our results on everyday recordings.

Fitting this picture, Odinaka et al.~\cite{Odinaka2010} found EERs below 7\% when training and testing on data from a single session, yet much higher EERs (below 20\%) when training and testing across sessions with gaps of up to 6 months. In these sessions, data was recorded for five minutes in a seated position. Our study thus likely captured more challenging real life variability with regard to body postures and other contexts.

ECG biometrics on more long-term recordings are evaluated by Labati et al.~\cite{DonidaLabati2018}: They relied on data from a public database with 24h recordings to train a deep CNN which achieved an EER of 5.81\%. In contrast to our study, the dataset was recorded with a medical-grade device and improved with manually reviewed QRS/beat detection. Moreover, the authors reported to have discarded noisy recordings. Thus, our data likely contains higher noise and reflects practical applications without manual data review and using wearable non-medical sensors. While optimising a specific classifier was not at the focus of our work here, in future work we plan to apply and evaluate deeper models on our data as well, such as the one in this related work.

In conclusion, we see our results as a first indication of what to expect from ECG biometrics on data matching practical applications, that is, using non-medical wearable sensors in daily life. In particular, authentication performances obtained on lab or ``cleaned'' data might often have to be considered optimistic. Hence, we call for future interdisciplinary research to combine machine learning and HCI perspectives, to conduct further studies beyond the lab and move closer towards usable and reliable real-life applications. Examples include access control in special working areas with a high demand for security (e.g. laboratories, data centers, or power plants).

\subsection{Training ECG Biometrics Systems on Real World Data}

Comparing the different evaluation scenarios revealed that for training the classification system on real world data it is important to capture ECG over several days.

We attribute this to the variance of heartbeats which might vary between days or even shorter timeframes. This is in line with existing work~\cite{Lourenco2014, Odinaka2010, Pouryayevali2014}. In addition, in the wild we expect extra noise due to varying real life contexts, such as activities, body postures and movements, and so on. These factors likely also contribute to the need of more training data from multiple days (i.e. from more diverse contexts). 

To quantify this, training on two days instead of one decreased EER relative by 1.11\%; training on three days by 2.37\%. Thus, we recommend that real world systems should take into account new training data on a regular basis. 

\subsection{Practical Use of ECG Among Other Biometrics}
Compared to other biometrics, for example fingerprint, ECG is less accurate~\cite{Hsiao2016}. For a real world system we thus see ECG as an additional factor in a multi-modal setup. This has already been shown as promising in combination with fingerprint~\cite{Hsiao2016} and face recognition~\cite{Israel2004}. 

Moreover, ECG is often attributed to provide information about liveness of a person~\cite{Fatemian2010, Hejazi2017, Islam2015, Srivastva2018, Ye2011, Zhao2012}. However, we do not share the view that ECG inherently provides secure liveness detection, since related work has demonstrated a successful attack~\cite{Eberz2017}: Using a waveform generator, the authors tricked an ECG sensor and biometrics system into authenticating despite not being attached to a living human. However, a multi-modal system could benefit from ECG as a measure for continuous authentication, since it can be sensed continuously and not only at certain points in time, such as fingerprints. In this regard, we consider a combination of a multitude of biometric factors as more akin to human liveness detection (e.g. combining fingerprint, ECG, face recognition, and behavioural biometrics). Investigating how to realise such combinations in applications in a usable way presents plenty of opportunities for future research.

\subsection{Implications for Everyday ECG Trackers}

As we worked with ECG field recordings, our data was rather noisy. The noise might be caused by electrical interference, as well as body motions, loosely fitted chest straps, or dry or oily skin. Direct contact of the electrodes to the skin is crucial. We used non-medical grade tracking devices with dry electrodes, which might have influenced the signal quality as well. In contrast, a medical grade devices works with adhesive electrodes, multiple leads, and might have better electronics built in. At the time of our study, we consider the tracking devices and chest straps as state of the art in non-medical ECG tracking. Nevertheless, based on our study experiences and results we see opportunities to improve aspects of such wearable trackers: For example, a chest strap could embed an array of electrodes. The tracking device could then consider all these measures, or decide for the pair of electrodes with the best signal quality. In the latter case, the tracker would also need to evaluate signal quality. This could be done with an integrated quality model based on machine learning, in line with work by Yin et al.~\cite{Yin2017}, who introduced an embedded ECG biometrics architecture with a quality check. 

\subsection{Signal Characteristics \& Challenges in Everyday ECG}
While applying our digital signal processing, we found it especially challenging to detect the P and T waves of a cardiac cycle. Their amplitudes were difficult to separate from surrounding noise. Therefore, we decided to rely solely on the QRS complex for identification, as those peaks appeared as most stable. Based on our data, we thus see the need for future work towards real world systems to focus on improved fiducial extraction. Incorporating all prominent characteristics of a cardiac cycle, such as the P wave, T wave, in addition to the QRS complex, might improve overall performance since more features could be computed. This might provide a more complete picture about a person's heart activity.

Alternatively, frequency domain features could be used~\cite{Hsiao2016, Zhao2018}. Work related to research on heart rate variability already noted that spectral analysis of low frequency features can require relatively long recordings to output features of sufficient quality~\cite{Malik1996}. This might be contrary to the need of a short identification phase. However, frequency features could be applied in a continuous authentication scenario. 

\subsection{Reflections on Privacy and Wider Implications}

In general, we showed that distinguishing people via ECG biometrics in everyday life is feasible. The sensing technology used in our study is already available on the market. We believe it is thus important for research on this topic to reflect on wider implications, beyond the technical evaluations. 

As a basic starting point for this reflection, consider that over 60\% of our participants after the experience deemed it as (very) unlikely that they would want to wear an ECG tracker for a biometric security application. This could mean that people would not like to buy and wear a device merely for ECG biometrics, which would motivate integration of ECG trackers into widespread devices such as smartphones or watches. This was already done in case of the Apple Watch 4 but with the main goal of health monitoring, not security. Alternatively, our participants' responses could reflect a general aversion against biometric applications.

For biometrics, we see a main potential concern in what might happen with the raw data and where it is stored. For instance, in commercial contexts, employers might store ECG data for identifying authorised personnel, but this data implicitly also contains information on people's health and lifestyle. Even in our study experience we saw a link of this kind: We excluded one person because of health conditions which would have likely affected the ECG data. 
More generally, research revealed that heart rate variability adapts to respiration~\cite{Aysin2006, Malik1996}, circadian rhythm~\cite{Massin2000}, exposure to daytime noise~\cite{Pitz2013}, alcohol consumption~\cite{Wang2018}, and mental stress~\cite{Kim2018}. 

This should be considered in the light of enterprise authentication based on ECG as a realistic use case. For example, companies offer wristbands which can capture fingerprints plus ECG for authentication\footnote{Nymi: \url{https://nymi.com/}}. It remains unclear whether such systems analyse data directly on the device, or if raw data is stored and processed on remote servers. We see this as a crucial difference from a privacy-sensitive perspective. Another approach might be the computation of cryptographic keys based on ECG~\cite{Karimian2017}. 

In past incidents, biometric databases have been the target of attacks, for example leaking fingerprint data~\cite{Guardian2015}, plus facial recognition data~\cite{Taylor2019}. Affected users might not be able to use their fingerprints for authentication anymore.

For ECG, it might be possible to recover, since it is less permanent and changes over time~\cite{Labati2013} in contrast to fingerprints~\cite{Cole2006}. However, leaked ECG data risks revealing peoples' health states and lifestyle choices.
In conclusion, we see privacy as one of the most important considerations when designing such biometric systems with physiological sensors. 

\section{Conclusion}

So far, ECG biometrics studies have been carried out mostly in the lab under controlled conditions and using medical-grade sensors which might not adequately reflect the conditions of practical real-world applications. Hence, existing work called for field studies on ECG biometrics~\cite{Singh2012a}. To address this gap, we presented an in-depth study of ECG biometrics with a wearable non-medical sensor in the wild. 

Overall, we found that it is generally possible to distinguish people via ECG data recorded in everyday life. Making decisions on 15 heartbeats, and training on data across a week, we found a best EER of 9.15\%. This is overall comparable to findings by lab studies. However, further analyses revealed that more realistic evaluation assumptions from an HCI and user perspective are much more challenging, for example expecting a system to only ask for user enrollment efforts on the first day of deployment. While we showed that results can be improved by making decisions on more data and data from multiple days, our study thus indicates that future research should aim to evaluate ECG biometrics systems not (only) in the lab but in the intended application contexts.

Our study also contributes to the bigger picture that more and more biometric modalities are viable to be measured in daily life. A broader implication worth highlighting is that it becomes increasingly important not to adopt a simplified view on ``biometrics'' as a single concept. For instance, as shown with the examples in our privacy discussion, the challenges and risks implied by using fingerprints vs ECG biometrics are considerably different. Working with everyday data, such as in our study here, thus highlights the need for responsible application design for real world biometrics applications.

\section{Acknowledgements}

We would like to thank movisens GmbH for providing us with the ECGmove3 sensors. 
This project is funded by the Bavarian State Ministry of Science and the Arts in the framework of the Centre Digitisation.Bavaria (ZD.B).

%
%
%
%
%

\balance{}

\bibliographystyle{SIGCHI-Reference-Format}
\bibliography{bib}


\begin{thebibliography}{00}


\ifx \showCODEN    \undefined \def \showCODEN     #1{\unskip}     \fi
\ifx \showDOI      \undefined \def \showDOI       #1{{\tt DOI:}\penalty0{#1}\ }
  \fi
\ifx \showISBNx    \undefined \def \showISBNx     #1{\unskip}     \fi
\ifx \showISBNxiii \undefined \def \showISBNxiii  #1{\unskip}     \fi
\ifx \showISSN     \undefined \def \showISSN      #1{\unskip}     \fi
\ifx \showLCCN     \undefined \def \showLCCN      #1{\unskip}     \fi
\ifx \shownote     \undefined \def \shownote      #1{#1}          \fi
\ifx \showarticletitle \undefined \def \showarticletitle #1{#1}   \fi
\ifx \showURL      \undefined \def \showURL       #1{#1}          \fi

\bibitem{Akhter2016}
{Nazneen Akhter}, {Sumegh Tharewal}, {Vijay Kale}, {Ashish Bhalerao}, {and}
  {K.~V. Kale}. 2016.
\newblock \showarticletitle{{Heart-Based Biometrics and Possible Use of Heart
  Rate Variability in Biometric Recognition Systems}}.
\newblock In {\em Advances in Intelligent Systems and Computing}. Vol. 395.
  Springer, New Delhi, 15--29.
\newblock
\showISBNx{978-81-322-2648-2}
\showISSN{21945357}
\showDOI{%
\url{http://dx.doi.org/10.1007/978-81-322-2650-5_2}}


\bibitem{Arteaga-falconi2015}
{Juan~Sebastian Arteaga-Falconi}, {Hussein {Al Osman}}, {and} {Abdulmotaleb {El
  Saddik}}. 2016.
\newblock \showarticletitle{{ECG Authentication for Mobile Devices}}.
\newblock {\em IEEE Transactions on Instrumentation and Measurement\/} {65}, 3
  (2016), 591--600.
\newblock
\showISSN{0018-9456}
\showDOI{%
\url{http://dx.doi.org/10.1109/TIM.2015.2503863}}


\bibitem{Aysin2006}
{Benhur Aysin} {and} {Elif Aysin}. 2006.
\newblock \showarticletitle{{Effect of Respiration in Heart Rate Variability
  (HRV) Analysis}}. In {\em 2006 International Conference of the IEEE
  Engineering in Medicine and Biology Society}. IEEE, 1776--1779.
\newblock
\showISBNx{1-4244-0032-5}
\showISSN{1557-170X}
\showDOI{%
\url{http://dx.doi.org/10.1109/IEMBS.2006.260773}}


\bibitem{Biel2001}
{L. Biel}, {O. Pettersson}, {L. Philipson}, {and} {P. Wide}. 2001.
\newblock \showarticletitle{{ECG analysis: a new approach in human
  identification}}.
\newblock {\em IEEE Transactions on Instrumentation and Measurement\/} {50}, 3
  (jun 2001), 808--812.
\newblock
\showISBNx{0-7803-5276-9}
\showISSN{00189456}
\showDOI{%
\url{http://dx.doi.org/10.1109/19.930458}}


\bibitem{Bowyer2008}
{Kevin~W. Bowyer}, {Karen Hollingsworth}, {and} {Patrick~J. Flynn}. 2008.
\newblock \showarticletitle{{Image understanding for iris biometrics: A
  survey}}.
\newblock {\em Computer Vision and Image Understanding\/} {110}, 2 (may 2008),
  281--307.
\newblock
\showISSN{10773142}
\showDOI{%
\url{http://dx.doi.org/10.1016/j.cviu.2007.08.005}}


\bibitem{Buschek2018}
{Daniel Buschek}, {Benjamin Bisinger}, {and} {Florian Alt}. 2018.
\newblock \showarticletitle{{ResearchIME}}. In {\em Proceedings of the 2018 CHI
  Conference on Human Factors in Computing Systems - CHI '18}. ACM Press, New
  York, New York, USA, 1--14.
\newblock
\showISBNx{9781450356206}
\showDOI{%
\url{http://dx.doi.org/10.1145/3173574.3173829}}


\bibitem{Buschek2015chi}
{Daniel Buschek}, {Alexander {De Luca}}, {and} {Florian Alt}. 2015.
\newblock \showarticletitle{{Improving Accuracy, Applicability and Usability of
  Keystroke Biometrics on Mobile Touchscreen Devices}}. In {\em Proceedings of
  the 33rd Annual ACM Conference on Human Factors in Computing Systems - CHI
  '15}. ACM Press, New York, New York, USA, 1393--1402.
\newblock
\showISBNx{9781450331456}
\showDOI{%
\url{http://dx.doi.org/10.1145/2702123.2702252}}


\bibitem{Campbell1997}
{J.P. Campbell}. 1997.
\newblock \showarticletitle{{Speaker recognition: a tutorial}}.
\newblock {\it Proc. IEEE} {85}, 9 (1997), 1437--1462.
\newblock
\showISSN{00189219}
\showDOI{%
\url{http://dx.doi.org/10.1109/5.628714}}


\bibitem{Carreiras2016}
{Carlos Carreiras}, {Andr{\'{e}} Louren{\c{c}}o}, {Hugo Silva}, {and} {Ana
  Fred}. 2016.
\newblock \showarticletitle{{Evaluating Template Uniqueness in ECG
  Biometrics}}.
\newblock   {370} (2016).
\newblock
\showISBNx{978-3-319-26451-6}
\showDOI{%
\url{http://dx.doi.org/10.1007/978-3-319-26453-0}}


\bibitem{Chen2017}
{Ying Chen} {and} {Wenxi Chen}. 2017.
\newblock \showarticletitle{{Finger ECG-based authentication for healthcare
  data security using artificial neural network}}. In {\em 2017 IEEE 19th
  International Conference on e-Health Networking, Applications and Services
  (Healthcom)}. IEEE, 1--6.
\newblock
\showISBNx{978-1-5090-6704-6}
\showDOI{%
\url{http://dx.doi.org/10.1109/HealthCom.2017.8210804}}


\bibitem{Choi2016a}
{Hyun-Soo Choi}, {Byunghan Lee}, {and} {Sungroh Yoon}. 2016.
\newblock \showarticletitle{{Biometric Authentication Using Noisy
  Electrocardiograms Acquired by Mobile Sensors}}.
\newblock {\em IEEE Access\/}  {4} (2016), 1266--1273.
\newblock
\showISBNx{2169-3536 VO - 4}
\showISSN{2169-3536}
\showDOI{%
\url{http://dx.doi.org/10.1109/ACCESS.2016.2548519}}


\bibitem{Christov2004}
{Ivaylo~I. Christov}. 2004.
\newblock \showarticletitle{{Real time electrocardiogram QRS detection using
  combined adaptive threshold}}.
\newblock {\em BioMedical Engineering Online\/}  {3} (2004), 1--9.
\newblock
\showISBNx{1475-925X (Electronic). 1475-925X (Linking)}
\showISSN{1475925X}
\showDOI{%
\url{http://dx.doi.org/10.1186/1475-925X-3-28}}


\bibitem{Hsiao2016}
{{Chun-Chieh Hsiao}}, {{Shei-Wei Wang}}, {Robert Lin}, {and} {{Ren-Guey Lee}}.
  2016.
\newblock \showarticletitle{{Multiple biometric authentication for personal
  identity using wearable device}}. In {\em 2016 IEEE International Conference
  on Systems, Man, and Cybernetics (SMC)}. IEEE, 673--678.
\newblock
\showISBNx{978-1-5090-1897-0}
\showDOI{%
\url{http://dx.doi.org/10.1109/SMC.2016.7844318}}


\bibitem{Cole2006}
{Simon~A Cole}. 2006.
\newblock \showarticletitle{{History of Fingerprint Pattern Recognition}}.
\newblock In {\em Automatic Fingerprint Recognition Systems}. Springer, New
  York, 1--25.
\newblock
\showDOI{%
\url{http://dx.doi.org/10.1007/0-387-21685-5_1}}


\bibitem{DaSilva2013}
{Hugo~Placido da Silva}, {Ana Fred}, {Andre Lourenco}, {and} {Anil~K. Jain}.
  2013.
\newblock \showarticletitle{{Finger ECG signal for user authentication:
  Usability and performance}}. In {\em 2013 IEEE Sixth International Conference
  on Biometrics: Theory, Applications and Systems (BTAS)}. IEEE, 1--8.
\newblock
\showISBNx{978-1-4799-0527-0}
\showDOI{%
\url{http://dx.doi.org/10.1109/BTAS.2013.6712689}}


\bibitem{Darwaish2014}
{Shah~Faisal Darwaish}, {Esmiralda Moradian}, {Tirdad Rahmani}, {and} {Martin
  Knauer}. 2014.
\newblock \showarticletitle{{Biometric Identification on Android Smartphones}}.
\newblock {\em Procedia Computer Science\/}  {35} (2014), 832--841.
\newblock
\showISSN{18770509}
\showDOI{%
\url{http://dx.doi.org/10.1016/j.procs.2014.08.250}}


\bibitem{DonidaLabati2018}
{Ruggero {Donida Labati}}, {Enrique Mu{\~{n}}oz}, {Vincenzo Piuri}, {Roberto
  Sassi}, {and} {Fabio Scotti}. 2019.
\newblock \showarticletitle{{Deep-ECG: Convolutional Neural Networks for ECG
  biometric recognition}}.
\newblock {\em Pattern Recognition Letters\/}  {126} (sep 2019), 78--85.
\newblock
\showISSN{01678655}
\showDOI{%
\url{http://dx.doi.org/10.1016/j.patrec.2018.03.028}}


\bibitem{Dubin2008}
{Dale Dubin}. 2000.
\newblock {\em {Rapid Interpretation of EKG's}}.
\newblock Cover Publishing Company. 368 pages.
\newblock
\showISBNx{0912912065}


\bibitem{Eberz2017}
{Simon Eberz}, {Nicola Paoletti}, {Marc Roeschlin}, {Andrea Patani}, {Marta
  Kwiatkowska}, {and} {Ivan Martinovic}. 2017.
\newblock \showarticletitle{{Broken Hearted: How To Attack ECG Biometrics}}. In
  {\em Proceedings 2017 Network and Distributed System Security Symposium}.
  Internet Society, Reston, VA.
\newblock
\showISBNx{1-891562-46-0}
\showDOI{%
\url{http://dx.doi.org/10.14722/ndss.2017.23408}}


\bibitem{Engelse1979}
{W.~A.~H. Engelse} {and} {C. Zeelenberg}. 1979.
\newblock \showarticletitle{{A single scan algorithm for QRS-detection and
  feature extraction}}.
\newblock {\em Computers in cardiology\/} {6}, 1979 (1979), 37--42.
\newblock


\bibitem{Fatemian2010}
{S.~Zahra Fatemian}, {Foteini Agrafioti}, {and} {Dimitrios Hatzinakos}. 2010.
\newblock \showarticletitle{{HeartID: Cardiac biometric recognition}}. In {\em
  2010 Fourth IEEE International Conference on Biometrics: Theory, Applications
  and Systems (BTAS)}. IEEE, 1--5.
\newblock
\showISBNx{978-1-4244-7581-0}
\showDOI{%
\url{http://dx.doi.org/10.1109/BTAS.2010.5634493}}


\bibitem{Goldberger2012}
{Ary~L Goldberger}, {Luis A~N Amaral}, {Leon Glass}, {Jeffrey~M Hausdorff},
  {Plamen~Ch Ivanov}, {Roger~G Mark}, {Joseph~E Mietus}, {George~B Moody},
  {Chung-kang Peng}, {and} {H~Eugene Stanley}. 2000.
\newblock \showarticletitle{{PhysioBank, PhysioToolkit, and PhysioNet}}.
\newblock {\em Circulation\/}  {101} (jun 2000).
\newblock
\showISSN{0009-7322}
\showDOI{%
\url{http://dx.doi.org/10.1161/01.CIR.101.23.e215}}


\bibitem{Goode2014}
{Alan Goode}. 2014.
\newblock \showarticletitle{{Bring your own finger - How mobile is bringing
  biometrics to consumers}}.
\newblock {\em Biometric Technology Today\/}  {2014} (2014), 5--9.
\newblock
\showISBNx{0969-4765}
\showISSN{09694765}
\showDOI{%
\url{http://dx.doi.org/10.1016/S0969-4765(14)70088-8}}


\bibitem{Hamilton2002}
{P Hamilton}. 2002.
\newblock \showarticletitle{{Open source ECG analysis}}. In {\em Computers in
  Cardiology}. IEEE, 101--104.
\newblock
\showISBNx{0-7803-7735-4}
\showDOI{%
\url{http://dx.doi.org/10.1109/CIC.2002.1166717}}


\bibitem{Hejazi2017}
{Maryamsadat Hejazi}, {S.~A.~R. Al-Haddad}, {Shaiful~Jahari Hashim}, {Ahmad
  Fazli~Abdul Aziz}, {and} {Yashwant~Prasad Singh}. 2017.
\newblock \showarticletitle{{Non-fiducial based ECG biometric authentication
  using one-class Support Vector Machine}}. In {\em 2017 Signal Processing:
  Algorithms, Architectures, Arrangements, and Applications (SPA)}. IEEE,
  190--194.
\newblock
\showISBNx{978-83-62065-30-1}
\showDOI{%
\url{http://dx.doi.org/10.23919/SPA.2017.8166862}}


\bibitem{Huang2018}
{Chenyu Huang}, {Huangxun Chen}, {Lin Yang}, {and} {Qian Zhang}. 2018.
\newblock \showarticletitle{{BreathLive: Liveness Detection for Heart Sound
  Authentication with Deep Breathing}}.
\newblock {\em Proceedings of the ACM on Interactive, Mobile, Wearable and
  Ubiquitous Technologies\/} {2}, 1 (2018), 1--25.
\newblock
\showISSN{24749567}
\showDOI{%
\url{http://dx.doi.org/10.1145/3191744}}


\bibitem{Guardian2015}
{Associated~Press in Washington}. 2015.
\newblock US government hack stole fingerprints of 5.6 million federal
  employees.
\newblock   (23 September 2015).
\newblock
\showURL{%
\url{https://www.theguardian.com/technology/2015/sep/23/us-government-hack-stole-fingerprints}}
\newblock
\shownote{(last accessed 20th, September 2019).}


\bibitem{Irvine2003}
{J.~M. Irvine}, {S.~A. Israel}, {M~D Wiederhold}, {and} {B.~K. Wiederhold}.
  2003.
\newblock \showarticletitle{{A New Biometric: Human Identification from
  Circulatory Function}}.
\newblock {\em Joint Statistical Meetings of the American Statistical
  Association, San Francisco\/} (2003), 1957--1963.
\newblock


\bibitem{Irvine2001}
{John~M. Irvine}, {Brenda~K. Wiederhold}, {Lauren~W. Gavshon}, {Steven Israel},
  {Shannon~B. McGehee}, {Rodney Meyer}, {and} {Mark~D. Wiederhold}. 2001.
\newblock \showarticletitle{{Heart Rate Variability: A New Biometric for Human
  Identification}}. In {\em Proceedings of the International Conference on
  Artificial Intelligence (IC-AI'2001)}, Vol.~3. CSREA Press, Las Vegas,
  1106--1112.
\newblock


\bibitem{Islam2015}
{Md~Saiful Islam}. 2015.
\newblock \showarticletitle{{Heartbeat Biometrics for Remote Authentication
  Using Sensor Embedded Computing Devices}}.
\newblock {\em International Journal of Distributed Sensor Networks\/} {11}, 6
  (jun 2015).
\newblock
\showISSN{1550-1477}
\showDOI{%
\url{http://dx.doi.org/10.1155/2015/549134}}


\bibitem{Israel2004}
{S.A. Israel}, {W.T. Scruggs}, {W.J. Worek}, {and} {J.M. Irvine}. 2004.
\newblock \showarticletitle{{Fusing face and ECG for personal identification}}.
  In {\em 32nd Applied Imagery Pattern Recognition Workshop, 2003.
  Proceedings.} IEEE, 226--231.
\newblock
\showISBNx{0-7695-2029-4}
\showISSN{21642516}
\showDOI{%
\url{http://dx.doi.org/10.1109/AIPR.2003.1284276}}


\bibitem{Israel2005}
{Steven~A. Israel}, {John~M. Irvine}, {Andrew Cheng}, {Mark~D. Wiederhold},
  {and} {Brenda~K. Wiederhold}. 2005.
\newblock \showarticletitle{{ECG to identify individuals}}.
\newblock {\em Pattern Recognition\/} {38}, 1 (jan 2005), 133--142.
\newblock
\showISSN{00313203}
\showDOI{%
\url{http://dx.doi.org/10.1016/j.patcog.2004.05.014}}


\bibitem{Kang2016}
{Shin~Jae Kang}, {Seung~Yong Lee}, {Hyo~Il Cho}, {and} {Hyunggon Park}. 2016.
\newblock \showarticletitle{{ECG Authentication System Design Based on Signal
  Analysis in Mobile and Wearable Devices}}.
\newblock {\em IEEE Signal Processing Letters\/} {23}, 6 (2016), 805--808.
\newblock
\showISBNx{1070-9908 VO - 23}
\showISSN{10709908}
\showDOI{%
\url{http://dx.doi.org/10.1109/LSP.2016.2531996}}


\bibitem{Karimian2017}
{Nima Karimian}, {Zimu Guo}, {Mark Tehranipoor}, {and} {Domenic Forte}. 2017.
\newblock \showarticletitle{{Highly Reliable Key Generation From
  Electrocardiogram (ECG)}}.
\newblock {\em IEEE Transactions on Biomedical Engineering\/} {64}, 6 (2017),
  1400--1411.
\newblock
\showISBNx{0018-9294 VO - 64}
\showISSN{15582531}
\showDOI{%
\url{http://dx.doi.org/10.1109/TBME.2016.2607020}}


\bibitem{Khan2015}
{Salman~H. Khan}, {M. {Ali Akbar}}, {Farrukh Shahzad}, {Mudassar Farooq}, {and}
  {Zeashan Khan}. 2015.
\newblock \showarticletitle{{Secure biometric template generation for
  multi-factor authentication}}.
\newblock {\em Pattern Recognition\/} {48}, 2 (feb 2015), 458--472.
\newblock
\showISSN{00313203}
\showDOI{%
\url{http://dx.doi.org/10.1016/j.patcog.2014.08.024}}


\bibitem{Kim2018}
{Hye-Geum Kim}, {Eun-Jin Cheon}, {Dai-Seg Bai}, {Young~Hwan Lee}, {and}
  {Bon-Hoon Koo}. 2018.
\newblock \showarticletitle{{Stress and Heart Rate Variability: A Meta-Analysis
  and Review of the Literature}}.
\newblock {\em Psychiatry Investigation\/} {15}, 3 (mar 2018), 235--245.
\newblock
\showISBNx{1096-1186 (Electronic) 1043-6618 (Linking)}
\showISSN{1738-3684}
\showDOI{%
\url{http://dx.doi.org/10.30773/pi.2017.08.17}}


\bibitem{Pitz2013}
{Ute Kraus}, {Alexandra Schneider}, {Susanne Breitner}, {Regina Hampel},
  {Regina R{\"{u}}ckerl}, {Mike Pitz}, {Uta Geruschkat}, {Petra Belcredi},
  {Katja Radon}, {and} {Annette Peters}. 2013.
\newblock \showarticletitle{{Individual Daytime Noise Exposure during Routine
  Activities and Heart Rate Variability in Adults: A Repeated Measures Study}}.
\newblock {\em Environmental Health Perspectives\/} {121}, 5 (may 2013),
  607--612.
\newblock
\showISSN{0091-6765}
\showDOI{%
\url{http://dx.doi.org/10.1289/ehp.1205606}}


\bibitem{Kyoso2001}
{Masaki Kyoso} {and} {Akihiko Uchiyama}. 2001.
\newblock \showarticletitle{{Development of an ECG identification system}}. In
  {\em 2001 Conference Proceedings of the 23rd Annual International Conference
  of the IEEE Engineering in Medicine and Biology Society}, Vol.~4. IEEE,
  3721--3723.
\newblock
\showISBNx{0-7803-7211-5}
\showDOI{%
\url{http://dx.doi.org/10.1109/IEMBS.2001.1019645}}


\bibitem{Labati2013}
{Ruggero~Donida Labati}, {Roberto Sassi}, {and} {Fabio Scotti}. 2013.
\newblock \showarticletitle{{ECG biometric recognition: Permanence analysis of
  QRS signals for 24 hours continuous authentication}}. In {\em 2013 IEEE
  International Workshop on Information Forensics and Security (WIFS)}. IEEE,
  31--36.
\newblock
\showISBNx{978-1-4673-5593-3}
\showISSN{2157-4766}
\showDOI{%
\url{http://dx.doi.org/10.1109/WIFS.2013.6707790}}


\bibitem{Lourenco2014}
{Andre Lourenco}, {Carlos Carreiras}, {Hugo Silva}, {and} {Ana Fred}. 2014.
\newblock \showarticletitle{{ECG biometrics: A template selection approach}}.
  In {\em 2014 IEEE International Symposium on Medical Measurements and
  Applications (MeMeA)}. IEEE, 1--6.
\newblock
\showISBNx{978-1-4799-2921-4}
\showDOI{%
\url{http://dx.doi.org/10.1109/MeMeA.2014.6860081}}


\bibitem{Lourenco2011b}
{Andr{\'{e}} Louren{\c{c}}o}, {Hugo Silva}, {and} {Ana Fred}. 2011.
\newblock \showarticletitle{{Unveiling the Biometric Potential of Finger-Based
  ECG Signals}}.
\newblock {\em Computational Intelligence and Neuroscience\/}  {2011} (2011),
  1--8.
\newblock
\showISSN{1687-5265}
\showDOI{%
\url{http://dx.doi.org/10.1155/2011/720971}}


\bibitem{Lourenco2012}
{Andre Lourenco}, {Hugo Silva}, {and} {Ana Fred}. 2012.
\newblock \showarticletitle{{ECG-based biometrics: A real time classification
  approach}}. In {\em 2012 IEEE International Workshop on Machine Learning for
  Signal Processing}. IEEE, 1--6.
\newblock
\showISBNx{978-1-4673-1026-0}
\showDOI{%
\url{http://dx.doi.org/10.1109/MLSP.2012.6349735}}


\bibitem{Malik1996}
{Marek Malik}. 1996.
\newblock \showarticletitle{{Heart Rate Variability.}}
\newblock {\em Annals of Noninvasive Electrocardiology\/} {1}, 2 (apr 1996),
  151--181.
\newblock
\showISBNx{1542-474X}
\showISSN{1082-720X}
\showDOI{%
\url{http://dx.doi.org/10.1111/j.1542-474X.1996.tb00275.x}}


\bibitem{Massin2000}
{M.~M. Massin}. 2000.
\newblock \showarticletitle{{Circadian rhythm of heart rate and heart rate
  variability}}.
\newblock {\em Archives of Disease in Childhood\/} {83}, 2 (aug 2000),
  179--182.
\newblock
\showISBNx{1468-2044 (Electronic) 0003-9888 (Linking)}
\showISSN{00039888}
\showDOI{%
\url{http://dx.doi.org/10.1136/adc.83.2.179}}


\bibitem{Miao2015}
{Fen Miao}, {Yayu Cheng}, {Yi He}, {Qingyun He}, {and} {Ye Li}. 2015.
\newblock \showarticletitle{{A Wearable Context-Aware ECG Monitoring System
  Integrated with Built-in Kinematic Sensors of the Smartphone}}.
\newblock {\em Sensors\/} {15}, 5 (may 2015), 11465--11484.
\newblock
\showISBNx{8675586392201}
\showISSN{1424-8220}
\showDOI{%
\url{http://dx.doi.org/10.3390/s150511465}}


\bibitem{Odinaka2010}
{Ikenna Odinaka}, {Po-Hsiang Lai}, {Alan~D. Kaplan}, {Joseph~A. O'Sullivan},
  {Erik~J. Sirevaag}, {Sean~D. Kristjansson}, {Amanda~K. Sheffield}, {and}
  {John~W. Rohrbaugh}. 2010.
\newblock \showarticletitle{{ECG biometrics: A robust short-time frequency
  analysis}}. In {\em 2010 IEEE International Workshop on Information Forensics
  and Security}. IEEE, 1--6.
\newblock
\showISBNx{978-1-4244-9078-3}
\showISSN{2157-4766}
\showDOI{%
\url{http://dx.doi.org/10.1109/WIFS.2010.5711466}}


\bibitem{Odinaka2012}
{Ikenna Odinaka}, {Po-Hsiang Lai}, {Alan~D. Kaplan}, {Joseph~A. O'Sullivan},
  {Erik~J. Sirevaag}, {and} {John~W. Rohrbaugh}. 2012.
\newblock \showarticletitle{{ECG Biometric Recognition: A Comparative
  Analysis}}.
\newblock {\em IEEE Transactions on Information Forensics and Security\/} {7},
  6 (dec 2012), 1812--1824.
\newblock
\showISBNx{9782940373482}
\showISSN{1556-6013}
\showDOI{%
\url{http://dx.doi.org/10.1109/TIFS.2012.2215324}}


\bibitem{Pan1985}
{Jiapu Pan} {and} {Willis~J. Tompkins}. 1985.
\newblock \showarticletitle{{A Real-Time QRS Detection Algorithm}}.
\newblock {\em IEEE Transactions on Biomedical Engineering\/} {BME-32}, 3 (mar
  1985), 230--236.
\newblock
\showISBNx{0-7803-3811-1}
\showISSN{0018-9294}
\showDOI{%
\url{http://dx.doi.org/10.1109/TBME.1985.325532}}


\bibitem{Patro2017}
{Kiran~Kumar Patro} {and} {P.~Rajesh Kumar}. 2017.
\newblock \showarticletitle{{Effective Feature Extraction of ECG for Biometric
  Application}}.
\newblock {\em Procedia Computer Science\/}  {115} (2017), 296--306.
\newblock
\showISBNx{9441029815}
\showISSN{18770509}
\showDOI{%
\url{http://dx.doi.org/10.1016/j.procs.2017.09.138}}


\bibitem{Pfeuffer2019}
{Ken Pfeuffer}, {Matthias~J. Geiger}, {Sarah Prange}, {Lukas Mecke}, {Daniel
  Buschek}, {and} {Florian Alt}. 2019.
\newblock \showarticletitle{{Behavioural Biometrics in VR}}. In {\em
  Proceedings of the 2019 CHI Conference on Human Factors in Computing Systems
  - CHI '19}. ACM Press, New York, New York, USA, 1--12.
\newblock
\showISBNx{9781450359702}
\showDOI{%
\url{http://dx.doi.org/10.1145/3290605.3300340}}


\bibitem{Pouryayevali2014}
{Shahrzad Pouryayevali}, {Saeid Wahabi}, {Siddarth Hari}, {and} {Dimitrios
  Hatzinakos}. 2014.
\newblock \showarticletitle{{On establishing evaluation standards for ECG
  biometrics}}. In {\em 2014 IEEE International Conference on Acoustics, Speech
  and Signal Processing (ICASSP)}. IEEE, 3774--3778.
\newblock
\showISBNx{978-1-4799-2893-4}
\showISSN{15206149}
\showDOI{%
\url{http://dx.doi.org/10.1109/ICASSP.2014.6854307}}


\bibitem{Schuckers2002}
{Stephanie~A.C. Schuckers}. 2002.
\newblock \showarticletitle{{Spoofing and Anti-Spoofing Measures}}.
\newblock {\em Information Security Technical Report\/} {7}, 4 (dec 2002),
  56--62.
\newblock
\showISSN{13634127}
\showDOI{%
\url{http://dx.doi.org/10.1016/S1363-4127(02)00407-7}}


\bibitem{Seepers2017}
{Robert~M. Seepers}, {Christos Strydis}, {Ioannis Sourdis}, {and} {Chris~I. {De
  Zeeuw}}. 2017.
\newblock \showarticletitle{{Enhancing Heart-Beat-Based Security for mHealth
  Applications}}.
\newblock {\em IEEE Journal of Biomedical and Health Informatics\/} {21}, 1
  (jan 2017), 254--262.
\newblock
\showISBNx{2168-2208 (Electronic)$\backslash$r2168-2194 (Linking)}
\showISSN{2168-2194}
\showDOI{%
\url{http://dx.doi.org/10.1109/JBHI.2015.2496151}}


\bibitem{Shen2002}
{T.W. Shen}, {W.J. Tompkins}, {and} {Y.H. Hu}. 2002.
\newblock \showarticletitle{{One-lead ECG for identity verification}}. In {\em
  Proceedings of the Second Joint 24th Annual Conference and the Annual Fall
  Meeting of the Biomedical Engineering Society] [Engineering in Medicine and
  Biology}, Vol.~1. IEEE, 62--63.
\newblock
\showISBNx{0-7803-7612-9}
\showISSN{1094-687X}
\showDOI{%
\url{http://dx.doi.org/10.1109/IEMBS.2002.1134388}}


\bibitem{Singh2012a}
{Yogendra~Narain Singh} {and} {S~K Singh}. 2012.
\newblock \showarticletitle{{Evaluation of Electrocardiogram for Biometric
  Authentication}}.
\newblock {\em Journal of Information Security\/} {3}, 1 (2012), 39--48.
\newblock
\showISSN{2153-1234}
\showDOI{%
\url{http://dx.doi.org/10.4236/jis.2012.31005}}


\bibitem{Sprager2017}
{Sebastijan Sprager}, {Roman Trobec}, {and} {Matjaz~B. Juric}. 2017.
\newblock \showarticletitle{{Feasibility of biometric authentication using
  wearable ECG body sensor based on higher-order statistics}}. In {\em 2017
  40th International Convention on Information and Communication Technology,
  Electronics and Microelectronics (MIPRO)}. IEEE, 264--269.
\newblock
\showISBNx{978-953-233-090-8}
\showDOI{%
\url{http://dx.doi.org/10.23919/MIPRO.2017.7973431}}


\bibitem{Sriram2009}
{Janani~C. Sriram}, {Minho Shin}, {Tanzeem Choudhury}, {and} {David Kotz}.
  2009.
\newblock \showarticletitle{{Activity-aware ECG-based patient authentication
  for remote health monitoring}}. In {\em Proceedings of the 2009 international
  conference on Multimodal interfaces - ICMI-MLMI '09}. ACM Press, New York,
  New York, USA, 297--304.
\newblock
\showISBNx{9781605587721}
\showISSN{1557-170X}
\showDOI{%
\url{http://dx.doi.org/10.1145/1647314.1647378}}


\bibitem{Srivastva2018}
{Ranjeet Srivastva} {and} {Yogendra~Narain Singh}. 2018.
\newblock \showarticletitle{{Identifying Individuals Using Fourier and
  Discriminant Analysis of Electrocardiogram}}.
\newblock In {\em International Conference on Mathematics and Computing}.
  Vol.~1. Springer, Singapore, 286--295.
\newblock
\showISBNx{9789811300233}
\showISSN{18650929}
\showDOI{%
\url{http://dx.doi.org/10.1007/978-981-13-0023-3_27}}


\bibitem{Tan2017}
{Robin Tan} {and} {Marek Perkowski}. 2017.
\newblock \showarticletitle{{Toward Improving Electrocardiogram (ECG) Biometric
  Verification using Mobile Sensors: A Two-Stage Classifier Approach}}.
\newblock {\em Sensors\/} {17}, 2 (feb 2017), 410.
\newblock
\showISSN{1424-8220}
\showDOI{%
\url{http://dx.doi.org/10.3390/s17020410}}


\bibitem{Tantawi2015}
{Manal~M. Tantawi}, {Kenneth Revett}, {Abdel-Badeeh Salem}, {and} {Mohamed~F.
  Tolba}. 2015.
\newblock \showarticletitle{{A wavelet feature extraction method for
  electrocardiogram (ECG)-based biometric recognition}}.
\newblock {\em Signal, Image and Video Processing\/} {9}, 6 (sep 2015),
  1271--1280.
\newblock
\showISSN{1863-1703}
\showDOI{%
\url{http://dx.doi.org/10.1007/s11760-013-0568-5}}


\bibitem{Taylor2019}
{Josh Taylor}. 2019.
\newblock Major breach found in biometrics system used by banks, UK police and
  defence firms.
\newblock   (14 August 2019).
\newblock
\showURL{%
\url{https://www.theguardian.com/technology/2019/aug/14/major-breach-found-in-biometrics-system-used-by-banks-uk-police-and-defence-firms}}
\newblock
\shownote{(last accessed 20th, September 2019).}


\bibitem{Tolosana2018}
{Ruben Tolosana}, {Ruben Vera-Rodriguez}, {Julian Fierrez}, {and} {Javier
  Ortega-Garcia}. 2018.
\newblock \showarticletitle{{Exploring Recurrent Neural Networks for On-Line
  Handwritten Signature Biometrics}}.
\newblock {\em IEEE Access\/}  {6} (2018), 5128--5138.
\newblock
\showISSN{2169-3536}
\showDOI{%
\url{http://dx.doi.org/10.1109/ACCESS.2018.2793966}}


\bibitem{Wang2018}
{Wen-Fong Wang}, {Ching-Yu Yang}, {and} {Yan-Fu Wu}. 2018.
\newblock \showarticletitle{{SVM-based classification method to identify
  alcohol consumption using ECG and PPG monitoring}}.
\newblock {\em Personal and Ubiquitous Computing\/} {22}, 2 (apr 2018),
  275--287.
\newblock
\showISBNx{0077901710420}
\showISSN{1617-4909}
\showDOI{%
\url{http://dx.doi.org/10.1007/s00779-017-1042-0}}


\bibitem{Ye2011}
{Can Ye}, {B.~V. K.~Vijaya Kumar}, {and} {Miguel~Tavares Coimbra}. 2011.
\newblock \showarticletitle{{Human identification based on ECG signals from
  wearable health monitoring devices}}. In {\em Proceedings of the 4th
  International Symposium on Applied Sciences in Biomedical and Communication
  Technologies - ISABEL '11}. ACM Press, New York, New York, USA, 1--5.
\newblock
\showISBNx{9781450309134}
\showDOI{%
\url{http://dx.doi.org/10.1145/2093698.2093723}}


\bibitem{Yin2017}
{Shihui Yin}, {Minkyu Kim}, {Deepak Kadetotad}, {Yang Liu}, {Chisung Bae},
  {Sang~Joon Kim}, {Yu Cao}, {and} {Jae-sun Seo}. 2017.
\newblock \showarticletitle{{A 1.06 $\mu$w smart ecg processor in 65 nm cmos
  for real-time biometric authentication and personal cardiac monitoring}}. In
  {\em 2017 Symposium on VLSI Circuits}. IEEE.
\newblock
\showISBNx{978-4-86348-614-0}
\showDOI{%
\url{http://dx.doi.org/10.23919/VLSIC.2017.8008563}}


\bibitem{Zhang2016}
{Liping Zhang}, {Kai Xing}, {Zhonghu Xu}, {Junbo Wang}, {Shuo Zhang}, {and}
  {Jing Xu}. 2016.
\newblock \showarticletitle{{Human Recognizer: An ECG based Live biometric
  Fingerprint}}. In {\em Proceedings of the 1st ACM Workshop on Privacy-Aware
  Mobile Computing - PAMCO '16}. ACM Press, New York, New York, USA, 21--27.
\newblock
\showISBNx{9781450343466}
\showDOI{%
\url{http://dx.doi.org/10.1145/2940343.2940347}}


\bibitem{Zhao2012}
{Chen~Xing Zhao}, {Tom Wysocki}, {Foteini Agrafioti}, {and} {Dimitrios
  Hatzinakos}. 2012.
\newblock \showarticletitle{{Securing handheld devices and fingerprint readers
  with ECG biometrics}}. In {\em 2012 IEEE Fifth International Conference on
  Biometrics: Theory, Applications and Systems (BTAS)}. IEEE, 150--155.
\newblock
\showISBNx{978-1-4673-1385-8}
\showDOI{%
\url{http://dx.doi.org/10.1109/BTAS.2012.6374570}}


\bibitem{Zhao2018}
{Zhidong Zhao}, {Yefei Zhang}, {Yanjun Deng}, {and} {Xiaohong Zhang}. 2018.
\newblock \showarticletitle{{ECG authentication system design incorporating a
  convolutional neural network and generalized S-Transformation}}.
\newblock {\em Computers in Biology and Medicine\/}  {102} (nov 2018),
  168--179.
\newblock
\showISSN{00104825}
\showDOI{%
\url{http://dx.doi.org/10.1016/j.compbiomed.2018.09.027}}


\bibitem{Zhou2011}
{Yingbo Zhou} {and} {Ajay Kumar}. 2011.
\newblock \showarticletitle{{Human Identification Using Palm-Vein Images}}.
\newblock {\em IEEE Transactions on Information Forensics and Security\/} {6},
  4 (dec 2011), 1259--1274.
\newblock
\showISSN{1556-6013}
\showDOI{%
\url{http://dx.doi.org/10.1109/TIFS.2011.2158423}}


\end{thebibliography}

\end{document}